\pdfoutput=1
\documentclass[aps,prapplied,superscriptaddress,twocolumn,floatfix,10pt]{revtex4-2}

\usepackage{amsmath}
\usepackage{iftex}
\ifPDFTeX
  \usepackage[utf8]{inputenc}
  \usepackage[T1]{fontenc}
  \usepackage{newtxtext}
\else
  \usepackage{fontspec}
\fi
\usepackage{newtxmath}

\usepackage{graphicx}
\usepackage[dvipsnames]{xcolor}
\usepackage[normalem]{ulem}
\usepackage{chemformula}
\usepackage{placeins,etoolbox}
\preto\subsection{\FloatBarrier}

\usepackage[colorlinks=true,citecolor=blue,linkcolor=blue,urlcolor=blue]{hyperref}
\setcitestyle{super}

\graphicspath{{./media/}}

\begin{document}

\title{Precursor stoichiometry controls dopant structure, defect chemistry, and valence in Mn-doped \texorpdfstring{\ch{ZnGa2O4}}{ZnGa2O4} nanocrystals}

\author{M\'aty\'as~M.~Rudolf}
\affiliation{HUN-REN Wigner Research Centre for Physics, Institute for Solid State Physics and Optics, P.O.\ Box 49, H-1525 Budapest, Hungary}
\affiliation{Department of Physical Chemistry and Materials Science, Faculty of Chemical Technology and Biotechnology, Budapest University of Technology and Economics, M\H{u}egyetem rakpart 3, H-1111 Budapest, Hungary}

\author{G\'abor~Bortel}
\affiliation{HUN-REN Wigner Research Centre for Physics, Institute for Solid State Physics and Optics, P.O.\ Box 49, H-1525 Budapest, Hungary}

\author{Nikoletta~Jegeny\'es}
\affiliation{HUN-REN Wigner Research Centre for Physics, Institute for Solid State Physics and Optics, P.O.\ Box 49, H-1525 Budapest, Hungary}

\author{\'Arp\'ad~Jakab}
\affiliation{HUN-REN Wigner Research Centre for Physics, Institute for Solid State Physics and Optics, P.O.\ Box 49, H-1525 Budapest, Hungary}

\author{L\'aszl\'o~Bencs}
\affiliation{HUN-REN Wigner Research Centre for Physics, Institute for Solid State Physics and Optics, P.O.\ Box 49, H-1525 Budapest, Hungary}

\author{Ferenc~Simon}
\affiliation{Stavropoulos Center for Complex Quantum Matter, Department of Physics and Astronomy, University of Notre Dame, Notre Dame, Indiana 46556, USA}
\affiliation{HUN-REN Wigner Research Centre for Physics, Institute for Solid State Physics and Optics, P.O.\ Box 49, H-1525 Budapest, Hungary}
\affiliation{Department of Physics, Institute of Physics, Budapest University of Technology and Economics, M\H{u}egyetem rakpart 3, H-1111 Budapest, Hungary}

\author{Adam~Gali}
\affiliation{HUN-REN Wigner Research Centre for Physics, Institute for Solid State Physics and Optics, P.O.\ Box 49, H-1525 Budapest, Hungary}
\affiliation{Department of Atomic Physics, Institute of Physics, Budapest University of Technology and Economics, M\H{u}egyetem rakpart 3, H-1111 Budapest, Hungary}
\affiliation{MTA-WFK Lend\"ulet ``Momentum'' Semiconductor Nanostructures Research Group, P.O.\ Box 49, H-1525 Budapest, Hungary}

\author{D\'avid~Beke}
\email{beke.david@wigner.hun-ren.hu}
\affiliation{HUN-REN Wigner Research Centre for Physics, Institute for Solid State Physics and Optics, P.O.\ Box 49, H-1525 Budapest, Hungary}
\affiliation{Kand\'o K\'alm\'an Faculty of Electrical Engineering, \'Obuda University, Tavaszmez\H{o} 17, H-1084 Budapest, Hungary}

\begin{abstract}
The luminescence of doped nanocrystals is usually optimised through the activator; here we show that the host cation ratio is an equally decisive, and largely overlooked, design variable. Mn-doped \ch{ZnGa2O4} nanocrystals (Mn:Zn $= 4\times10^{-4}$) were grown hydrothermally at nominal Zn/Ga ratios between $0.30$ and $0.75$ (stoichiometric: $0.50$) and post-annealed at $400$--$1100~^{\circ}$C, and the resulting composition--temperature matrix was mapped by X-ray- and UV-excited luminescence, electron paramagnetic resonance (EPR), X-ray diffraction with Rietveld analysis, Raman spectroscopy and electron microscopy. The as-grown crystals are phase-pure spinel at every composition and accommodate their off-stoichiometry internally as a point-defect inventory, reflected in a $0.4\%$ variation of the lattice parameter; on annealing, the excess gallium is exsolved as $\beta$-\ch{Ga2O3} in amounts set by the solution composition. The \ch{Mn^2+} emission separates into two optima, near-stoichiometric crystals annealed at $400$--$600~^{\circ}$C and Zn-excess crystals annealed at $1100~^{\circ}$C, the latter being twice as bright under X-ray excitation. Under X-ray and $365$~nm excitation the two optima are separated by a minimum at $800~^{\circ}$C that coincides with a collapse of the \ch{Mn^2+} EPR population and has no counterpart in the Raman linewidths, identifying transient, defect-compensated oxidation of the dopant rather than lattice degradation as its origin. The most Zn-rich composition shows neither \ch{Mn^2+} nor \ch{Mn^4+} emission at any annealing stage, whereas its trace \ch{Cr^3+} emission is strong, indicating that manganese is rejected already during growth. Solution stoichiometry therefore acts on two levels, dopant capture and native-defect inventory, while annealing controls dopant valence and lattice order. The brightest high-temperature phosphor is thus not the nominally stoichiometric one, but the composition whose excess is consumed on its way to stoichiometry.
\end{abstract}


\maketitle

\section{Introduction}
\label{sec:intro}

Zinc gallate (\ch{ZnGa2O4}, ZGO) is a normal spinel with a band gap of $4.4$--$5.2$~eV \cite{Sampath1998} that combines chemical and thermal robustness with two crystallographically distinct cation sites, tetrahedral \ch{Zn^2+} (A) and octahedral \ch{Ga^3+} (B). A single host can therefore accommodate activators of different valence and coordination preference, and because the lattice converts ionising radiation into visible light efficiently, ZGO has been developed for a wide range of optical applications: as a low-voltage cathodoluminescent phosphor for field-emission and vacuum-fluorescent displays \cite{Kim2005,Zhang2014}, as a transparent ultra-wide-band-gap semiconductor \cite{Galazka2019}, and, when activated with \ch{Cr^3+}, as a near-infrared persistent phosphor for \emph{in vivo} imaging \cite{Bessiere2014,Maldiney2014} and as an X-ray-excited nanotransducer \cite{Beke2021,Rudolf2022} for X-ray-activated photodynamic therapy \cite{Kamkaew2016,Song2018,Chen2017} and its combination with immunotherapy \cite{LuK2018}. Tetrahedral \ch{Mn^2+} provides a narrow green emission band near $505$~nm that is attractive for displays and scintillation \cite{Shea1994,Zhang2014,Wada2011}, and Mn-activated spinels are being explored for dosimetry \cite{Honjo2025,Takebuchi2025}, for multi-band and afterglow emission \cite{Kazarinov2025,Yin2026,Si2021}, and as X-ray-chargeable light sources for deep-tissue therapy, either releasing therapeutic gases \cite{Xue2020} or, in related Mn-activated oxides, activating photosensitisers \cite{Liu2022,Kamkaew2016}.

In all of these applications the optical response is governed by the native-defect chemistry of the host. Undoped ZGO is itself a self-activated blue phosphor whose emission arises from donor--acceptor recombination within the Ga--O sublattice \cite{Jeong1998,Ren2014,Wada2011}, and nominally stoichiometric ZGO contains a few percent of $\mathrm{Ga}_{\mathrm{Zn}}^{\bullet}$ and $\mathrm{Zn}_{\mathrm{Ga}}'$ antisite pairs \cite{Huang2023,DeVos2016,Yang2023}. This intrinsic disorder is also the origin of the persistent luminescence of ZGO:\ch{Cr^3+}. \ch{Cr^3+} ions perturbed by a neighbouring antisite pair give rise to the N$_2$ zero-phonon line, and the local electric field of the same defects separates the electron--hole pairs created by visible-light excitation of \ch{Cr^3+} and stores the carriers on the antisite defects, from which their thermally assisted recombination at \ch{Cr^3+} sustains the emission for hours \cite{Bessiere2014,Gourier2014,Basavaraju2015,DeVos2016}. This afterglow, chargeable with visible light or X-rays, is the basis of autofluorescence-free \emph{in vivo} imaging \cite{leMasnedeChermont2007,Pan2012,Maldiney2014}, of rechargeable optical probes and sensors \cite{Li2015,Sun2021}, and of anti-counterfeiting labels \cite{Si2021}, and a weaker, vacancy-related afterglow is known for ZGO:\ch{Mn^2+} as well \cite{Uheda1997,Luchechko2019}. Control over the antisite and vacancy population therefore controls the optical properties \cite{Li2016,Yang2023}, which can lead to brighter emission and better-controlled persistent luminescence for Mn activators.

Among the synthesis parameters, the cation stoichiometry governs this population most directly, and its leverage is well documented for \ch{Cr^3+}-activated gallates. Zinc deficiency in ZGO:\ch{Cr^3+} is accommodated by antisite formation and deepens the trap distribution \cite{Zhuang2013,Bessiere2011}; aliovalent co-substitution of the octahedral site, as in \ch{ZnGa_{2-$x$}(Mg/Ge)_{$x$}O4}, creates charged defect clusters around the emitting ion and enhances both the steady-state and the persistent emission \cite{Zhu2021,Yang2023}; Zn deficiency in mixed aluminogallate spinels tunes cation inversion and the local distortion of the \ch{Cr^3+} environment \cite{Lim2025}; and deliberately unbalanced cation ratios in \ch{ZnAl2O4} redistribute activators and trap states \cite{Ratio2025}. For manganese, host composition and synthesis atmosphere together determine whether the tetrahedral or octahedral position is occupied as \ch{Mn^2+} or higher valences \cite{Song2019,Yin2026}. However, annealing temperature also controls both the inversion degree and the dopant valence \cite{Cornu2014,Dwibedi2018}. In most of these studies, stoichiometry and thermal treatment are varied separately, and the materials are prepared by solid-state reaction above $1000~^{\circ}$C, a temperature at which cation interdiffusion and \ch{ZnO} volatilisation have already partly erased the precursor ratio \cite{Galazka2019}. In solution-phase growth, i.e.\ under hydrothermal conditions, the cation ratio of the medium is often far from the final cation ratio, and whether a trace dopant enters the lattice is governed by surface chemistry and site competition during growth~\cite{Erwin2005,Norris2008}. For hydrothermal ZGO, the solution Zn/Ga ratio is known to control phase purity, particle size, lattice parameter, and composition \cite{Hirano2000,Chen2006,Li2015,Zhang2014}, and the reagent ratio has recently also been identified as a key parameter for the brightness of solvothermal ZGO:\ch{Cr^3+} nanoparticles \cite{Drozdowski2025}. We have shown that the \ch{Cr^3+} emission of hydrothermal ZGO nanocrystals is strongly modulated by the annealing temperature \cite{Rudolf2022}. Consequently, the brightest hydrothermally synthesised phosphors are rarely those synthesised at the stoichiometric ratio, and whether post-annealing affects the optimum remains unstudied.

To close this gap, we map manganese-activated ZGO nanocrystals across the precursor-ratio and annealing-temperature plane. A series of ZGO:Mn nanocrystals was grown hydrothermally at nominal Zn/Ga ratios between $0.30$ and $0.75$ with a fixed Mn concentration (Mn:Ga $= 2\times10^{-4}$, i.e.\ Mn:Zn $= 2.7$--$6.7\times10^{-4}$ across the series), and each sample was annealed between $400$ and $1100~^{\circ}$C. X-ray-excited optical luminescence (XEOL), photoluminescence (PL) and photoluminescence-excitation spectroscopy (PLE), electron paramagnetic resonance (EPR) spectroscopy, powder X-ray diffraction (XRD) with Rietveld analysis, Raman spectroscopy, and scanning electron microscopy with energy-dispersive X-ray spectroscopy (SEM-EDS) were recorded over the full matrix. Manganese was chosen because \ch{Mn^2+} substitutes isovalently for tetrahedral \ch{Zn^2+} and needs no charge compensation; its spin-forbidden $^{4}T_{1}(^{4}G)\rightarrow{}^{6}A_{1}(^{6}S)$ emission at $\approx505$~nm is sensitive to the local coordination \cite{Zhang2014,Wada2011}; and its ground-state population can be counted, because the $^{55}$Mn hyperfine sextet makes EPR a quantitative probe of isolated \ch{Mn^2+}. \ch{Mn^3+} ($3d^{4}$, $S = 2$) is silent at X-band, so that oxidation appears as a loss of signal, whereas \ch{Mn^4+} ($3d^{3}$) has its own red $^{2}E\rightarrow{}^{4}A_{2}$ emission and EPR signature \cite{Cornu2014,Vikas2023}. At $\approx130$~ppm of the dopant, concentration quenching and Mn--Mn exchange are negligible, and changes in emission or EPR intensity reflect the host rather than the dopant ensemble.

We show that hydrothermal synthesis followed by annealing yields Mn-doped \ch{ZnGa2O4} single-crystalline nanoparticles whose \ch{Mn^2+} emission is about an order of magnitude more intense than that of samples prepared by solid-state reaction, while the emission of the trace \ch{Cr^3+} impurity is relatively weaker (Fig.~\ref{fig:HTvsSS}). We also show that the precursor ratio during hydrothermal synthesis governs both the dopant and host properties, and that annealing affects defect chemistry. The as-grown crystals are single-phase, well-faceted nanocrystals at every composition and store their off-stoichiometry as antisite and vacancy defects; annealing exsolves this excess as $\beta$-\ch{Ga2O3} in amounts encoded by the precursor ratio. Interestingly, the \ch{Mn^2+} emission has two optima, near-stoichiometric precursor ratios annealed at $400~^{\circ}$C and Zn-excess precursor ratios annealed at $1100~^{\circ}$C, which under X-ray and $365$~nm excitation are divided by a minimum at $800~^{\circ}$C due to the formation of \ch{Mn^3+}. Annealing also modulates the intrinsic defect population and, therefore, the defect-related emission. These results translate into quantitative design rules for spinel phosphors and scintillators.

\section{Results and Discussion}
\label{sec:results}

The nanocrystals were synthesised hydrothermally at $230~^{\circ}$C for $13$~h following our previously reported procedure \cite{Beke2021,Rudolf2022}. Details are provided in Sec.~\ref{sec:methods}. Samples are labelled by the nominal Zn/Ga molar ratio of the precursor solution, $x$, normalised to Ga $= 2$ ($x = 0.30$--$0.75$; $x = 0.50$ is stoichiometric), and by the annealing temperature $T_{\rm A}$. ``As-prepared'' denotes the hydrothermal product ($230~^{\circ}$C, $13$~h) before any further treatment.

\subsection{Solution composition defines crystal shape and stoichiometry}
\label{sec:stoichiometry}

\begin{figure*}[htp]
    \includegraphics[width=\textwidth]{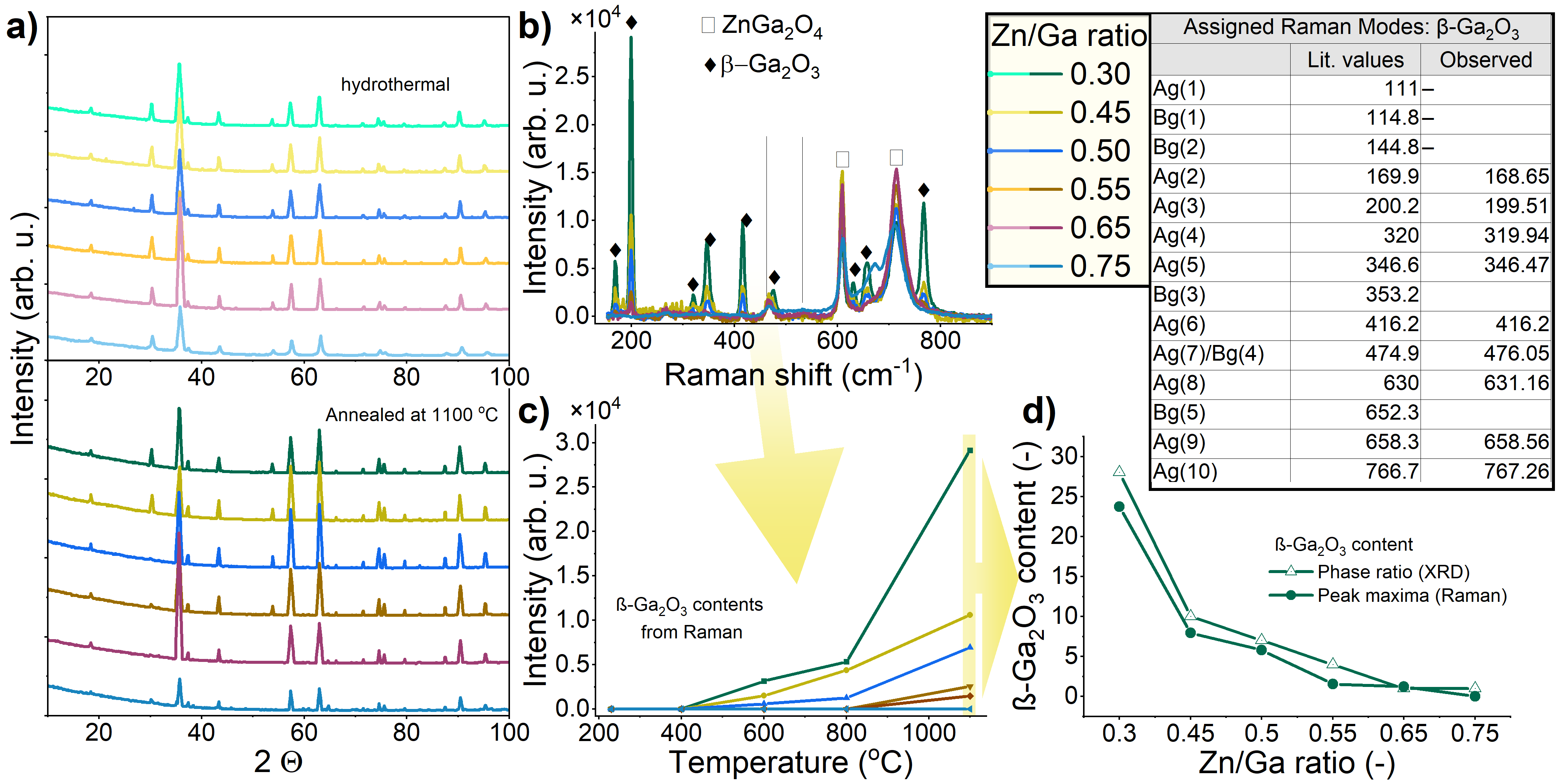}
    \caption{
        Phase evolution upon annealing. (A) XRD patterns of the as-prepared (top) and $1100~^{\circ}$C-annealed (bottom) samples; $\blacklozenge$ marks $\beta$-\ch{Ga2O3} reflections. (B) Raman spectra after $1100~^{\circ}$C annealing with the \ch{ZnGa2O4} ($\triangledown$) and $\beta$-\ch{Ga2O3} ($\blacklozenge$) modes labelled; the table compares measured and literature $\beta$-\ch{Ga2O3} $A_{g}$ Raman modes. (C) $\beta$-\ch{Ga2O3} Raman intensity versus $T_{\rm A}$ for each composition. (D) $\beta$-\ch{Ga2O3} content after $1100~^{\circ}$C from Rietveld phase fractions (wt\%) and from scaled Raman intensities, versus the nominal Zn/Ga ratio.}
    \label{fig:phase}
\end{figure*}

\begin{figure*}[htp]
    \includegraphics[width=\textwidth]{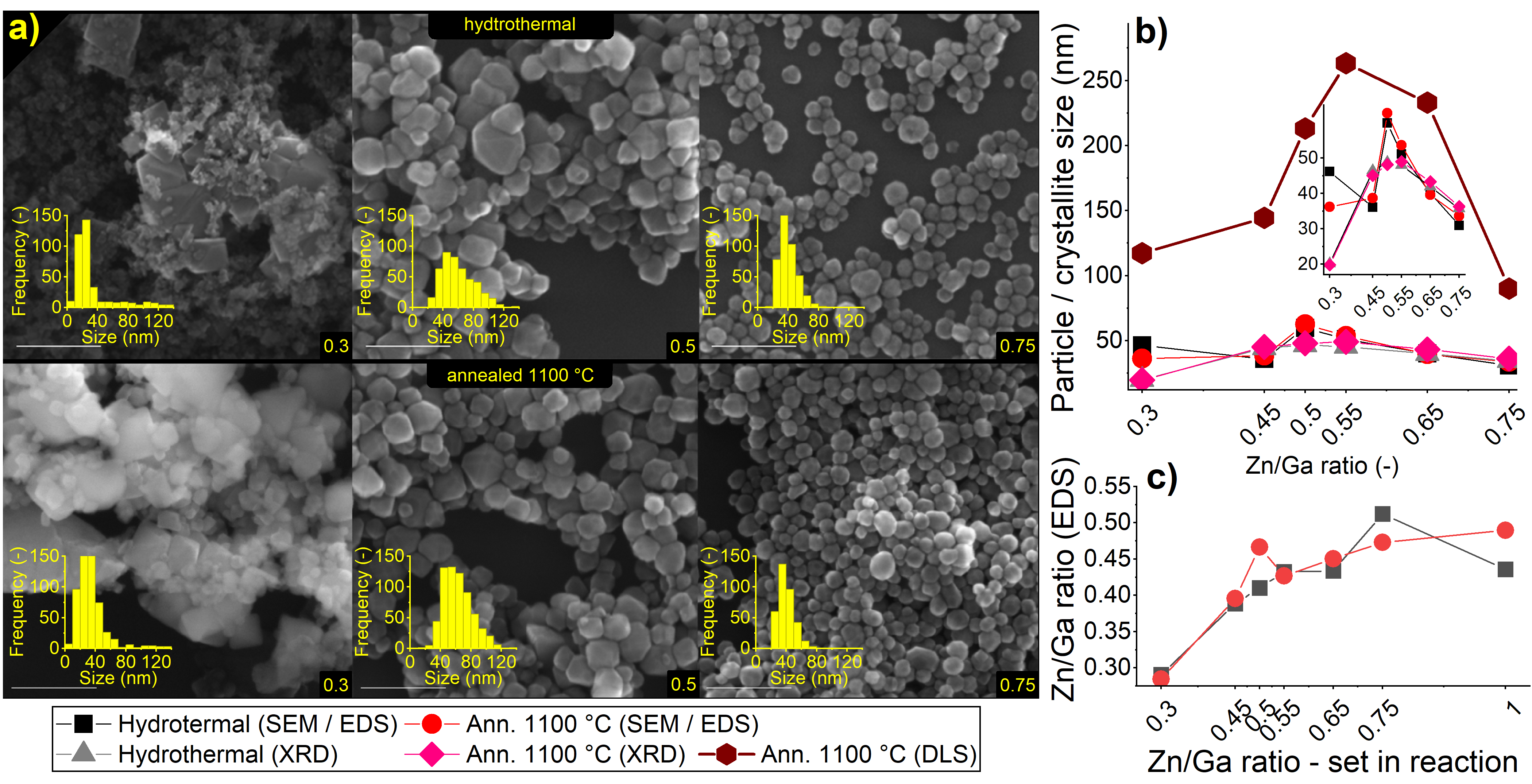}
    \caption{
        Morphology and composition of the as-prepared and annealed series. (A) SEM images of the as-prepared ($230~^{\circ}$C, top) and $1100~^{\circ}$C-annealed (bottom) samples with $x = 0.30$, $0.50$ and $0.65$; insets: particle-size histograms. (B) Mean particle size (SEM) and coherent-domain size (Rietveld) versus $x$ for $T_{\rm A} = 230$--$1100~^{\circ}$C. (C) Zn/Ga ratio measured by SEM-EDS versus the nominal ratio, for as-prepared and $1100~^{\circ}$C-annealed samples; the dashed line marks the stoichiometric value $0.5$.}
    \label{fig:morphology}
\end{figure*}

All members of the series crystallised directly under hydrothermal conditions as single-phase spinel [Fig.~\ref{fig:phase}(A)]. No \ch{GaO(OH)}, \ch{ZnO} or \ch{Ga2O3} reflections were detected at any composition, in contrast to lower-temperature hydrothermal routes in which \ch{GaO(OH)} forms as an intermediate \cite{Hirano2000,Chen2006}. The crystals are nevertheless Ga-rich, with off-stoichiometry decreasing as $x$ increases, and a Zn:Ga ratio of 1:2 is reached only when about twice the stoichiometric amount of Zn is supplied in solution \cite{Rudolf2022} [Fig.~\ref{fig:morphology}(C)], in agreement with the aqueous-synthesis optimum reported by Li \emph{et al.} \cite{Li2015}. Spot analyses give Zn/Ga $= 0.32$ in octahedron-rich and $0.28$ in octahedron-poor regions of the same sample, which lies within the uncertainty of the method and suggests that the octahedral shape formation is a kinetic response of the growing crystal to the Ga-rich medium and not the signature of a separate Ga-rich particle population.

In the as-prepared samples, the cubic lattice parameter increased monotonically from $a = 8.305$~\AA{} ($x = 0.30$) to $8.341$~\AA{} ($x = 0.75$; Fig.~\ref{fig:S26}), reaching the reference value of stoichiometric \ch{ZnGa2O4} \cite{ICDD860410}. The sign and magnitude of the deviation follow the Shannon radii \cite{Shannon1976}. The tetrahedral \ch{Zn^2+} ($r_{\rm IV} = 0.60$~\AA) is replaced by the smaller \ch{Ga^3+} ($r_{\rm IV} = 0.47$~\AA); the resulting $\mathrm{Ga}_{\mathrm{Zn}}^{\bullet}$ antisites, charge-compensated by zinc vacancies, contract the cell while preserving the $Fd\bar{3}m$ framework, as reported for Zn-deficient spinels \cite{Lim2025,Zhuang2013}.

The nominal Zn/Ga ratio nevertheless strongly affects the properties and composition and, to a lesser extent, the size distribution (Fig.~\ref{fig:morphology}). Ga-excess ($x = 0.30$ and $0.45$) yielded well-faceted octahedra. From the stoichiometric solution composition upwards, the octahedra gave way first to truncated polyhedra and finally to irregular polyhedral particles of $40$--$60$~nm. The octahedron is the $\{111\}$-faceted morphology of the spinel lattice and develops when the ratio of growth rates $R = v_{\langle100\rangle}/v_{\langle111\rangle}$ approaches $1.73$ \cite{Wang2000}; a Ga-rich medium therefore drives $R$ towards the octahedral limit, whereas growth under Zn excess is nearly isotropic. The mean particle size from SEM agrees with the coherent-domain size from XRD [Fig.~\ref{fig:morphology}(B); additional SEM images and EDS data in Fig.~\ref{fig:S3}], which, together with the well-defined facets, indicates that the particles are single crystals. Remarkably, the majority of the imaged nanoparticles were well-faceted single crystals, although no surfactant was used in the hydrothermal synthesis. To the best of our knowledge, octahedral ZGO crystals have so far been reported only with shape-directing surfactants such as CTAB \cite{Liu2013}, by chemical vapour deposition \cite{Zhang2008}, or as building blocks of spindle-shaped aggregates \cite{Yin2016,Chen2006}, which makes a surfactant-free route to this morphology a notable result of this work. Indeed, here the synthesis is additive-free, and the Zn/Ga ratio is the sole shape-control parameter. It is also worth mentioning that single crystals formed despite the non-stoichiometric crystal composition, and their shape was preserved under the applied heat treatment.

\subsection{Annealing converts dissolved off-stoichiometry into \texorpdfstring{$\beta$-\ch{Ga2O3}}{beta-Ga2O3}}
\label{sec:annealing}

Laboratory XRD cannot, however, resolve how this excess is distributed over the two cation sites. \ch{Zn^2+} and \ch{Ga^3+} differ by a single electron, and their X-ray scattering factors are nearly identical. We tested the sensitivity of the refinement by comparing twelve nested parametrisations in which the site occupancies, the inversion degree, a Zn$\leftrightarrow$Ga exchange, and the isotropic atomic displacement parameters $B_{\mathrm{eq}}$ were released in all combinations (see Sec.~\ref{sec:methods}). Releasing any cation-distribution parameter alone leaves the weighted profile residual unchanged ($R_{\mathrm{wp}} = 7.08$--$7.13\%$ against $7.13\%$ for the fixed model of the $x = 0.30$ as-prepared sample), and in the fully free model these parameters converge to scattered, partly bound-pinned values. Per-sample displacement parameters alone, in contrast, reproduce the entire improvement of the fully free model ($R_{\mathrm{wp}} = 3.94\%$ against $3.93\%$). Cation-distribution parameters are therefore statistically indeterminate on these data, and all structural values below refer to the fixed-distribution model (unit occupancies, zero inversion, refined $B_{\mathrm{eq}}$).

Within this model, $B_{\mathrm{eq}}$ is dominated by static, not thermal, displacements and serves as an effective measure of lattice disorder. It is largest in the as-prepared crystals ($B_{\mathrm{eq}} \approx 2$--$4$~\AA$^{2}$), decreases at $600~^{\circ}$C from $3.4$ to $1.9$~\AA$^{2}$ with increasing $x$, i.e.\ it scales with the gallium excess, and falls to $0.5$--$1.3$~\AA$^{2}$ after annealing at $1100~^{\circ}$C, the range expected for a well-ordered oxide at room temperature. The diffraction data are thus consistent with an ideal normal spinel after $1100~^{\circ}$C, but they remain blind to antisite disorder at the few-percent level.

Raman spectroscopy, in contrast, is sensitive to the cation distribution. For the normal spinel structure, group theory predicts five Raman-active zone-centre modes, $\Gamma_{\rm Raman} = A_{1g} + E_{g} + 3F_{2g}$ \cite{White1967,DIppolito2015}. In \ch{ZnGa2O4} these are observed at $\approx467$ ($F_{2g(2)}$), $611$ ($F_{2g(3)}$) and $714$~cm$^{-1}$ ($A_{1g}$), whereas the $E_{g}$ and $F_{2g(1)}$ modes, predicted near $370$--$395$ and $185$~cm$^{-1}$, carry very little intensity \cite{vanGorkom1973,Stokey2020}. The $A_{1g}$ mode is the symmetric breathing of the oxygen tetrahedron around the A-site cation, and, because the octahedral $16d$ site is a centre of inversion, the B-site cations do not move in any Raman-active mode. Cation inversion breaks the translational symmetry and activates additional bands \cite{Cynn1992,Lazzeri2006,DIppolito2015,LuX2018}. Indeed, in the as-prepared crystals the $F_{2g(3)}$ and $A_{1g}$ modes are broad and bridged by such a band at $\approx640$--$680$~cm$^{-1}$ (Fig.~\ref{fig:S5}), which loses most of its intensity at $600~^{\circ}$C and survives after $1100~^{\circ}$C only as a weak shoulder in the Zn-rich compositions [Fig.~\ref{fig:phase}(B)]. The Raman spectra thus confirm that the as-prepared crystals store their off-stoichiometry as antisite disorder, which annealing progressively removes.

Annealing releases the stored cation excess. At $T_{\rm A} = 600~^{\circ}$C the $a(x)$ curve flattens, and after $1100~^{\circ}$C all samples converge to $a = 8.333$--$8.337$~\AA, i.e.\ to (near-)stoichiometric \ch{ZnGa2O4}. The expelled gallium appears as $\beta$-\ch{Ga2O3} in both the Raman and the XRD data. Its $A_{g}$ phonons (cf.\ Refs.~\cite{Dohy1982,Machon2006}) emerge from $T_{\rm A} = 600~^{\circ}$C onward [Fig.~\ref{fig:phase}(B),(C)], and the corresponding reflections appear in the $1100~^{\circ}$C diffractograms [Fig.~\ref{fig:phase}(A)]. The Rietveld phase fraction after $1100~^{\circ}$C decreases monotonically with $x$, and the Raman peak intensities follow it closely [Fig.~\ref{fig:phase}(D)].

\subsection{Composition- and temperature-resolved luminescence}
\label{sec:spectroscopy}

The studied emission spectra contained three main components (Fig.~\ref{fig:S6}). Under both X-ray and UV excitation, the emission was dominated by the green $^{4}T_{1}(^{4}G)\rightarrow{}^{6}A_{1}(^{6}S)$ transition of tetrahedral \ch{Mn^2+} at $\approx510$~nm \cite{Zhang2014,Wada2011,Kim2005}. A weak low-energy shoulder indicates a minority of \ch{Mn^2+} in higher-coordinated, antisite-like environments, as resolved in the spectral decomposition (Fig.~\ref{fig:S12}) \cite{Cornu2014}. A broad band centred near $580$~nm accompanies the \ch{Mn^2+} line in the as-prepared and moderately annealed samples, and is attributed to the self-activated Ga--O emission \cite{Ren2014,Wada2011}. Finally, all samples showed a structured deep-red multiplet between $680$ and $730$~nm, with sharp emission lines at $\approx688$ and $\approx696$~nm and broader peaks at $704$--$715$~nm, characteristic of \ch{Cr^3+} zero-phonon lines and phonon sidebands present as an unavoidable trace impurity~\cite{Bessiere2014,DeVos2016,Huang2023}. Indeed, despite the clear Cr-related PL signal, the Cr content was below the detection limit of atomic absorption spectroscopy. Spectral decomposition of the red band under $365$~nm excitation requires an additional component, which we tentatively assign to octahedral \ch{Mn^4+} ($^{2}E\rightarrow{}^{4}A_{2}$; Fig.~\ref{fig:S2}), known to emit in the same window in the aluminate analogues \cite{Cornu2014,Vikas2023}.

\begin{figure}[htp]
    \includegraphics[width=\columnwidth]{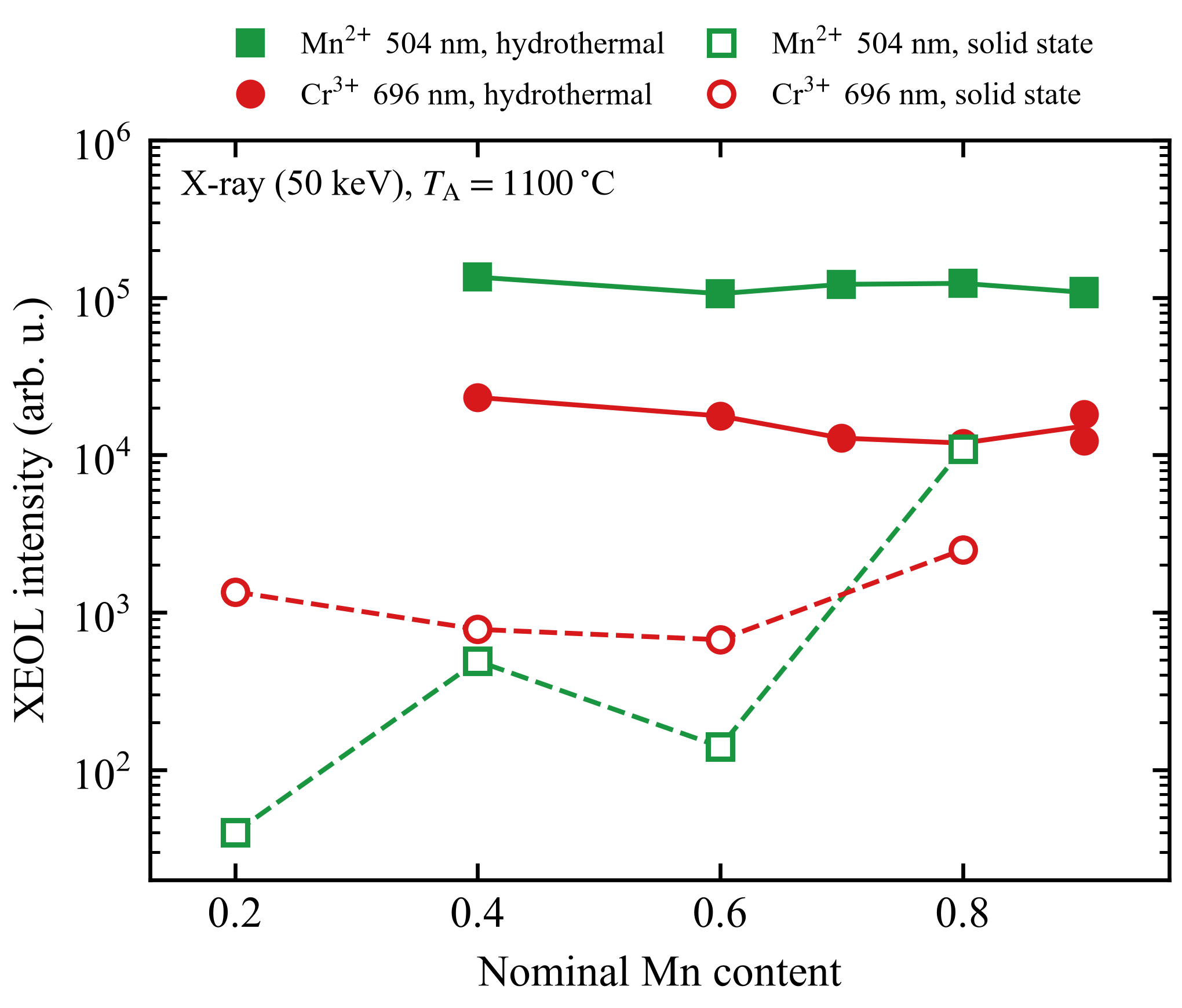}
    \caption{
        XEOL intensity of the \ch{Mn^2+} emission at $504$~nm (green squares) and of the \ch{Cr^3+} zero-phonon line at $696$~nm (red circles) for a hydrothermally prepared series (filled symbols, solid lines; $14$ samples) and a solid-state series (open symbols, dashed lines; $15$ samples), both annealed at $1100~^{\circ}$C and excited with $50$~keV X-rays, as a function of the nominal Mn content. Two hydrothermal samples were measured at $0.9$; the lines connect their mean. Note the logarithmic intensity scale.}
    \label{fig:HTvsSS}
\end{figure}

Figure~\ref{fig:HTvsSS} compares the X-ray-excited \ch{Mn^2+} ($504$~nm) and \ch{Cr^3+} ($696$~nm) emission intensities of two reference series annealed at $1100~^{\circ}$C, one prepared hydrothermally and one by direct solid-state reaction. The brightest hydrothermal sample emits about an order of magnitude more \ch{Mn^2+} light than the brightest solid-state sample ($1.35\times10^{5}$ versus $1.1\times10^{4}$~arb.~u.), and its \ch{Mn^2+}/\ch{Cr^3+} intensity ratio is also higher ($6$--$10$ versus at most $4.4$). The \ch{Mn^2+} XEOL intensity of the hydrothermal series varies by less than $30\%$ across the series, whereas in the solid-state series only the sample with the highest nominal loading ($0.8$) emits efficiently, and the other three remain below $5\%$ of its intensity.

\begin{figure*}[htp]
    \includegraphics[width=\textwidth]{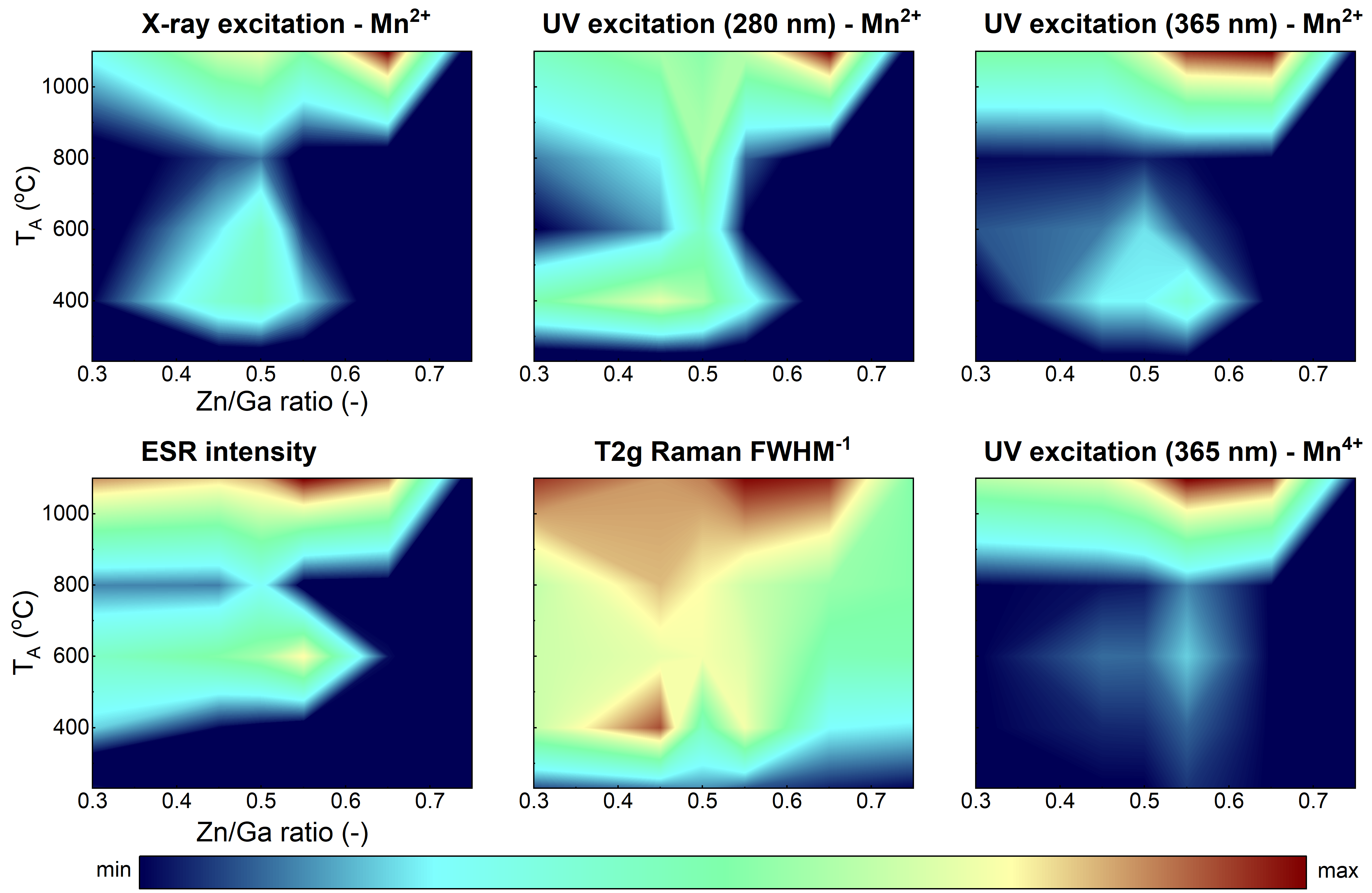}
    \caption{
        Composition--annealing maps of the \ch{Mn^2+} channel and of the structural probes. Top row: integrated intensity of the \ch{Mn^2+} emission under X-ray (A), $280$~nm (B) and $365$~nm (C) excitation. Bottom row: doubly integrated intensity of the \ch{Mn^2+} EPR sextet (D), sharpness ($1/\mathrm{FWHM}$) of the $F_{2g(3)}$ Raman mode at $\approx611$~cm$^{-1}$ (E) and integrated intensity of the red component assigned to \ch{Mn^4+} under $365$~nm excitation (F). Each panel is normalised to its own maximum and interpolated between the measured grid points ($x = 0.30$, $0.45$, $0.50$, $0.55$, $0.65$, $0.75$; $T_{\rm A} = 230$, $400$, $600$, $800$, $1100~^{\circ}$C); $T_{\rm A} = 230~^{\circ}$C denotes the as-prepared samples. The $580$~nm defect band and the \ch{Cr^3+} emission are shown in Fig.~\ref{fig:allin}. Note that the Raman sharpness (E) increases monotonically through $800~^{\circ}$C, whereas the EPR population (D) and the X-ray- and $365$~nm-excited emission (A, C) collapse there.}
    \label{fig:channels}
\end{figure*}

The hydrothermal step therefore substantially improves the optical properties of Mn-doped \ch{ZnGa2O4}. Mapping the optical response across the whole composition--annealing matrix, however, reveals that a much lower annealing temperature can yield comparable emission, and the full data set allows us to identify how, and why, $x$ and $T_{\rm A}$ control the luminescence. We represent these data as contour maps whose axes are the two synthesis parameters, $x$ and $T_{\rm A}$, and whose colour scale gives the measured intensity. Figure~\ref{fig:channels} maps the \ch{Mn^2+} channel over the composition--temperature plane for X-ray, $280$~nm and $365$~nm excitation and compares it with the \ch{Mn^2+} EPR intensity, the sharpness of the $F_{2g(3)}$ Raman mode and the \ch{Mn^4+}-assigned component; the $580$~nm band and the \ch{Cr^3+} emission are shown for all channels in Fig.~\ref{fig:allin}, and the \ch{Mn^2+}/\ch{Mn^4+} maps under $365$~nm excitation in Fig.~\ref{fig:S2}. The \ch{Mn^2+} emission showed two maxima under every excitation (see the line profiles in Fig.~\ref{fig:Sint}). The first was reached by near-stoichiometric and mildly Ga-rich crystals ($x = 0.45$--$0.50$) after annealing at $400~^{\circ}$C, where the XEOL intensity rose by two orders of magnitude above its as-prepared level. The second, global optimum belongs to Zn-excess crystals annealed at $1100~^{\circ}$C, which reach twice the intensity of the first ($x = 0.65$); an independently synthesised second series reproduces this composition profile (Fig.~\ref{fig:S7}). Between the two optima the channels diverge. Under X-ray and $365$~nm excitation the intensity decreases with increasing annealing temperature and reaches a global minimum at $800~^{\circ}$C for all compositions. Under $280$~nm excitation, in contrast, the minimum is shallower and lies at $600~^{\circ}$C, and the intensity has already recovered at $800~^{\circ}$C, where the stoichiometric sample reaches $65\%$ of the maximum.

The EPR intensities follow the topology of the optical maps with one systematic shift [Fig.~\ref{fig:channels}(D); spectra and peak intensities in Figs.~\ref{fig:S13} and \ref{fig:S14}]: along the temperature axis the \ch{Mn^2+} population peaks at $600~^{\circ}$C rather than at $400~^{\circ}$C, collapses at $800~^{\circ}$C and recovers at $1100~^{\circ}$C, where it depends only weakly on composition between $x = 0.30$ and $0.65$, and vanishes at $x = 0.75$. Since the double integral of the sextet is proportional to the number of resonating spins, and the manganese content of a sample is fixed by the synthesis, the variation across the annealing series must reflect a change of the manganese environment or valence rather than of its concentration. Two mechanisms can remove intensity from the sextet. A strongly distorted or strained environment broadens the hyperfine lines and their fine-structure satellites until they merge with the baseline; alternatively, a change in the oxidation state can produce an EPR-silent species. Of the three valence states, only \ch{Mn^3+} ($3d^{4}$, $S = 2$) is silent at X-band, because its large zero-field splitting places the transitions outside the accessible field range \cite{Mansoor2019}. The first mechanism must follow the lattice order and can only improve with annealing; the second can be non-monotonic. Discriminating between them therefore requires an independent measure of lattice order that is blind to the manganese valence.

The Raman linewidths provide such a measure, resolved by sublattice [Fig.~\ref{fig:channels}(E) and Fig.~\ref{fig:S1}]. The $A_{1g}$ mode is the symmetric breathing of the oxygen tetrahedron around the A-site cation, \ch{(Zn,Mn)O4}, and its width is therefore an order parameter of the sublattice that hosts the activator, whereas the $F_{2g(3)}$ mode involves oxygen displacements that modulate the octahedral Ga--O bonds, so that its width reflects the distribution of octahedral environments \cite{Mohacek2017,DIppolito2015}. Neither width depends on the manganese valence, and, unlike the absolute Raman intensities, the inverse width ($1/\mathrm{FWHM}$, ``sharpness'') needs no normalisation between samples. The two modes behave differently. The $A_{1g}$ sharpness is highest for the stoichiometric solution ($x = 0.50$) at every $T_{\rm A}$ below $1100~^{\circ}$C, increases with $T_{\rm A}$, with the largest step between $400$ and $600~^{\circ}$C, and shows no anomaly at $800~^{\circ}$C. The $F_{2g(3)}$ sharpness, in contrast, mirrors the \ch{Mn^2+} emission under UV excitation: at $400~^{\circ}$C it has two local maxima on either side of stoichiometry ($x = 0.45$ and $0.55$), like the $365$~nm-excited emission, and it increases monotonically towards $1100~^{\circ}$C without any decrease at $800~^{\circ}$C, like the $280$~nm-excited emission. The lattice therefore orders continuously through the $800~^{\circ}$C step at which the EPR sextet and the X-ray- and $365$~nm-excited emission collapse, which excludes the broadening mechanism and leaves a change of the manganese valence as the origin of the $800~^{\circ}$C minimum.

\subsection{Mechanism: capture during growth, activation by annealing}
\label{sec:mechanism}

The optical, magnetic and vibrational data can be combined into a consistent picture with four elements: (i) a transient, defect-compensated oxidation of the dopant around $800~^{\circ}$C; (ii) a transfer-limited emission at low annealing temperature; (iii) a capture-limited emission after recrystallisation at $1100~^{\circ}$C; and (iv) the rejection of manganese at the Zn-rich extreme.

\emph{(i) The $800~^{\circ}$C minimum is a valence change, not lattice degradation.} Annealing in air between $\approx500$ and $900~^{\circ}$C oxidises \ch{Mn^2+} towards \ch{Mn^3+} ($3d^{4}$), which is EPR-silent at X-band and non-luminescent in the visible \cite{Cornu2014,Mansoor2019}. \ch{Mn^3+} is a non-Kramers ion ($S = 2$) with a zero-field splitting beyond the X-band range \cite{Mansoor2019}, and its $^{5}T_{2}\rightarrow{}^{5}E$ emission is a weak, broad and strongly Stokes-shifted near-infrared band at $1.1$--$1.5$~\textmu m that is quenched at room temperature by Jahn--Teller coupling \cite{Kueck2001}. The temperature agrees with the stability range of \ch{Mn2O3} in air ($\approx530$--$940~^{\circ}$C), above which the equilibrium returns to lower mean valence \cite{Grundy2003}. A similar non-monotonic evolution of the \ch{Mn^2+} EPR signal between $700$ and $1100~^{\circ}$C has been reported for ZGO:Mn phosphors \cite{Kim2005}, and a \ch{Mn^3+}~$\rightarrow$~\ch{Mn^2+} recovery at high temperature for \ch{ZnAl2O4}:Mn \cite{Cornu2014}. Oxidation of a tetrahedral dopant requires charge compensation, for example $\mathrm{Mn}_{\mathrm{Zn}}^{\bullet}$ balanced by $\mathrm{V}_{\mathrm{Zn}}''$ or $\mathrm{Zn}_{\mathrm{Ga}}'$. The defect-rich lattice at $600$--$800~^{\circ}$C supplies such compensators in abundance, whereas the recrystallised lattice at $1100~^{\circ}$C does not and forces manganese back into the isovalent \ch{Mn^2+}($T_{\rm d}$) configuration. The structural data exclude the alternative. Both Raman modes sharpen monotonically with $T_{\rm A}$, in step with the decrease of $B_{\mathrm{eq}}$ in the Rietveld refinements, i.e.\ crystallinity improves and distortions are annealed out through the very step at which the EPR sextet and the X-ray- and $365$~nm-excited emission collapse. The EPR map is therefore the superposition of two processes, the monotonic improvement of the \ch{Mn^2+} environment with annealing, and the valence change, which empties it at $800~^{\circ}$C; together, they place its optimum at $600~^{\circ}$C.

The absence of an $800~^{\circ}$C minimum under $280$~nm excitation refines this picture. At $4.4$~eV the excitation lies below the host absorption edge ($245$--$255$~nm, Table~\ref{tab:S1}) but inside the charge-transfer band of manganese in \ch{ZnGa2O4}, which is centred near $287$~nm ($4.3$~eV) \cite{Schuyt2024} and appears as a shoulder at $275$--$320$~nm in the excitation spectrum of the green emission \cite{Luchechko2019}, consistent with our PLE results (Figs.~\ref{fig:S22}--\ref{fig:S25}). Since the ground-state \ch{Mn^2+} population counted by EPR is at its minimum at $800~^{\circ}$C, the undiminished $280$~nm-excited emission requires a route to excited \ch{Mn^2+} that does not start from ground-state \ch{Mn^2+}. We tentatively assign it to a charge-transfer excitation of \ch{Mn^3+}, in which an electron is transferred from the oxygen ligands or from a neighbouring compensating acceptor to the dopant and the resulting (\ch{Mn^2+})$^{*}$ decays radiatively before the charge separation is reversed; a metastable \ch{Mn^3+} state that recaptures an electron only slowly is well documented for \ch{ZnO} \cite{Ciechan2016}. Indeed, the $580$~nm defect band, i.e.\ the donor--acceptor recombination of the host \cite{Jeong1998,Ren2014}, is strongest under $280$~nm excitation at $600$--$800~^{\circ}$C and is absent after $1100~^{\circ}$C. X-rays, which create band-to-band carriers throughout the crystal, and $365$~nm photons, which reach only the spin-forbidden $d$--$d$ bands of \ch{Mn^2+} and the defect states, have no access to this channel, and their output follows the EPR population.

\emph{(ii) At low annealing temperature the emission is transfer-limited.} The PLE spectra of annealed near-stoichiometric samples are dominated by the host absorption edge at $245$--$255$~nm, with only weak direct $d$--$d$ bands at $360$--$450$~nm (Figs.~\ref{fig:S23}--\ref{fig:S25}), and the composition dependence is weaker under $365$~nm excitation. Between $400$ and $600~^{\circ}$C the lattice is still defect-rich. Distorted Ga--O polyhedra and defect-perturbed regions act as efficient absorbers and energy donors for \ch{Mn^2+}, as established for the \ch{Ga2O3}--\ch{ZnO} system, in which the \ch{Mn^2+} emission maximises together with the density of distorted Ga--O units on the Ga-rich side of stoichiometry \cite{Wada2011}. Consistently, the $580$~nm defect emission is strongest in the Ga-rich crystals at intermediate temperatures, and it is completely annealed out at $1100~^{\circ}$C (Figs.~\ref{fig:S8} and \ref{fig:S9}). The emission intensity per EPR-detected \ch{Mn^2+} ion is nevertheless several times higher at $400~^{\circ}$C than at $600~^{\circ}$C, which indicates that the defect channel begins to compete with the \ch{Mn^2+} emission once it is fully developed; the optical optimum therefore lies one step below the EPR optimum on the temperature axis. The two Raman peak maxima of the $F_{2g(3)}$ sharpness at $x = 0.45$ and $0.55$ coincide with those of the UV-excited emission at $400~^{\circ}$C, which identifies the octahedral Ga--O framework, the absorber of the host under UV excitation \cite{Jeong1998,Wada2011}, as the bottleneck of the sensitised channel: excitation absorbed by the Ga--O sublattice has to migrate to \ch{Mn^2+}, and the same octahedral disorder that broadens $F_{2g(3)}$ provides the competing non-radiative and defect-emission channels. The $A_{1g}$ sharpness, in contrast, peaks at $x = 0.50$, i.e.\ the tetrahedral sublattice that hosts the activator is most regular for the stoichiometric solution, even though the crystal it produces is Ga-rich.

\emph{(iii) At $1100~^{\circ}$C the emission is capture-limited and favours compositions that end up stoichiometric.} $1100~^{\circ}$C is the temperature of the solid-state synthesis of \ch{ZnGa2O4}, and the growth of the coherent domains to $100$--$270$~nm, the relaxation of the lattice parameter and the exsolution of $\beta$-\ch{Ga2O3} all show that the nanocrystals recrystallise at this stage. The brightest $1100~^{\circ}$C samples have post-anneal compositions closest to Zn/Ga $= 0.5$ and a minimal $\beta$-\ch{Ga2O3} fraction. The Zn surplus in the synthesis compensates both for the intrinsically Zn-deficient growth and for the Zn loss expected at $1100~^{\circ}$C from the volatility of \ch{ZnO} \cite{Galazka2019}. Ga-rich compositions, in contrast, carry a $\beta$-\ch{Ga2O3} fraction whose $4.6$--$4.9$~eV absorption competes with that of the spinel, and the spinel/\ch{Ga2O3} interface may act as a non-radiative sink. The different penetration depths of the two excitations make this especially visible. X-rays probe the entire particle volume and require efficient carrier transport to the activator, whereas UV photons are absorbed within the outer few tens of nanometres. At $1100~^{\circ}$C the most Ga-rich member ($x = 0.30$) retains about a third of the maximum XEOL intensity but is nearly dark under UV-C excitation (Fig.~\ref{fig:S20}), consistent with surface-segregated $\beta$-\ch{Ga2O3} intercepting the UV-C light, whose photon energy lies above the $\beta$-\ch{Ga2O3} absorption edge. The PLE spectra show the same hierarchy: all annealed samples absorb strongly at the band edge, which shifts from $250$ to $290$~nm at $1100~^{\circ}$C as a \ch{Mn^2+}/charge-transfer shoulder develops \cite{Zhang2014}, whereas the as-prepared Zn-rich samples are excitable only through defect bands at $355$--$430$~nm.

\emph{(iv) At the Zn-rich extreme, manganese is not captured.} The collapse at $x = 0.75$ reveals the second, independent role of the solution stoichiometry. Under X-ray and $280$~nm excitation, the \ch{Mn^2+} emission of this composition never exceeds $2\%$ of the respective maximum at any $T_{\rm A}$, its $365$~nm-excited green signal below $1100~^{\circ}$C is entirely accounted for by the $580$~nm band, and its \ch{Mn^2+} EPR sextet is more than an order of magnitude weaker than in the other samples at every annealing stage. The \ch{Mn^4+}-assigned emission is equally absent, whereas the \ch{Cr^3+} emission of the same samples is among the strongest of the series (Fig.~\ref{fig:channels}). The residual red emission is excited through defect bands at $\approx265$ and $\approx370$~nm rather than through the host edge. With a large \ch{Zn^2+} surplus competing for the same tetrahedral sites at a Mn:Zn supply of only $4\times10^{-4}$, the trace dopant is rejected by the growing crystal, in line with the self-purification of doped nanocrystals \cite{Erwin2005,Norris2008}, and remains in solution or on the particle surface. The same hierarchy explains why the hydrothermal route clearly outperforms direct solid-state reaction: in a fifteen-sample solid-state reference series synthesised at $1100~^{\circ}$C, the green emission remains at least ten times below that of the hydrothermal series (Fig.~\ref{fig:HTvsSS}). The hydrothermal synthesis delivers manganese into the lattice before annealing, whereas the solid-state route must achieve incorporation and ordering simultaneously by high-temperature diffusion.

\begin{figure}[htp]
    \includegraphics[width=\columnwidth]{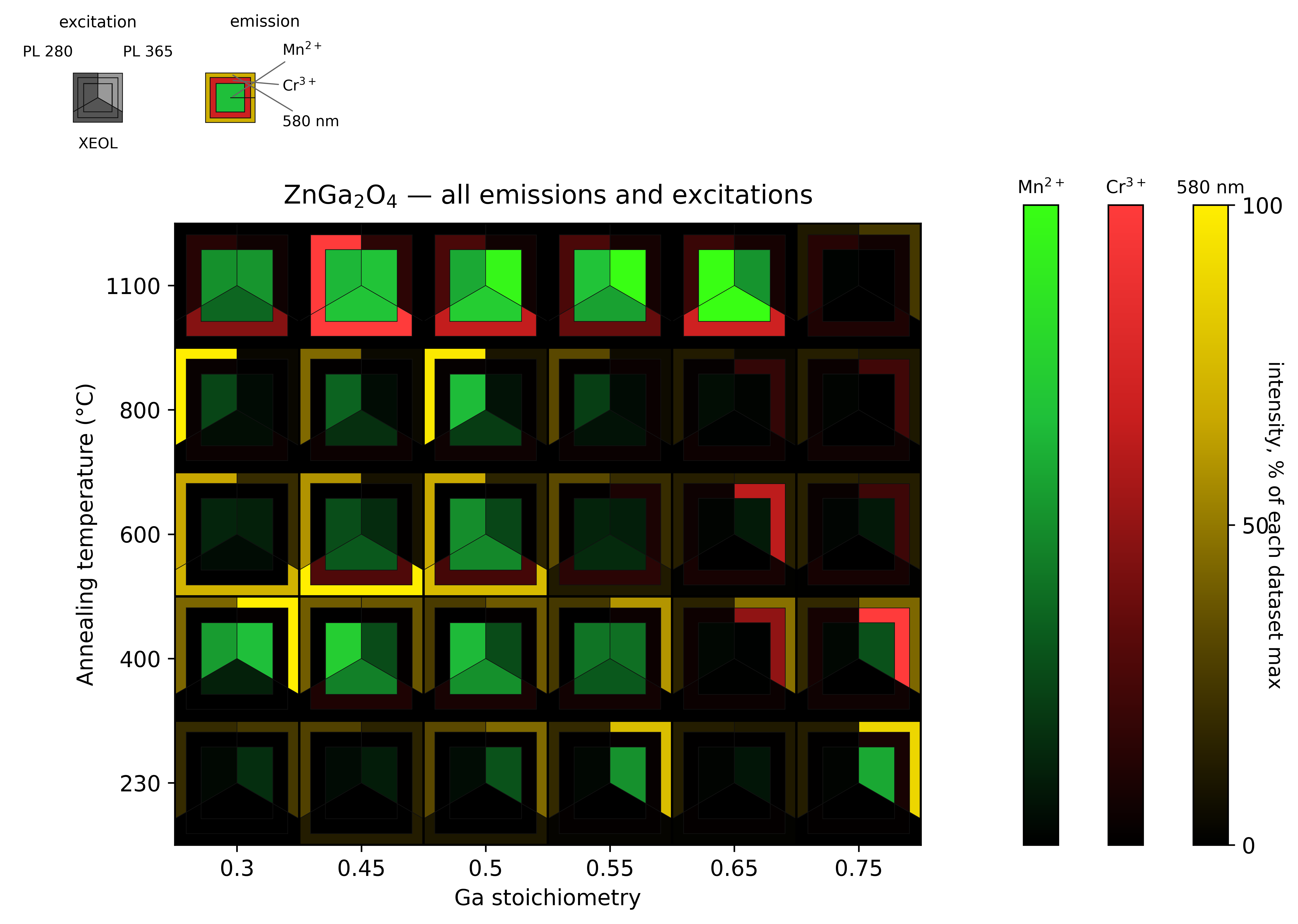}
    \caption{
        Emission intensity of the \ch{ZnGa2O4} sample series as a function of Ga stoichiometry (horizontal axis) and annealing temperature (vertical axis) for all nine excitation--emission channel combinations. Each cell corresponds to one sample and is divided into three sectors by excitation (upper left: PL under $280$~nm excitation; upper right: PL under $365$~nm excitation; lower half: X-ray-excited optical luminescence, XEOL; Mo-anode tube, $50$~kV) and into three concentric rings of equal area by emission channel (inner ring: \ch{Mn^2+} band, green scale; middle ring: \ch{Cr^3+} band, red scale; outer ring: $580$~nm band, yellow scale). Each of the nine channels is normalised to its own maximum, so that black denotes zero and full colour denotes the brightest sample of that channel; the maxima are $8.4\times10^{4}$, $5.9\times10^{4}$ and $1.3\times10^{5}$ counts for \ch{Mn^2+}, $1.5\times10^{5}$, $3.9\times10^{5}$ and $5.1\times10^{3}$ counts for \ch{Cr^3+}, and $1.2\times10^{4}$, $3.9\times10^{4}$ and $4.0\times10^{3}$ counts for the $580$~nm band under $280$~nm, $365$~nm and X-ray excitation, respectively. The two keys in the upper left define the sector and ring assignments. Note that the \ch{Mn^2+} and \ch{Cr^3+} emissions are brightest after annealing at $1100~^{\circ}$C for all but the Ga-richest sample, whereas the 580~nm band is confined to annealing temperatures of $800~^{\circ}$C and below, and that under 365~nm excitation the \ch{Cr^3+} emission dominates in the Ga-rich samples annealed at $400$--$800~^{\circ}$C.}
    \label{fig:allin}
\end{figure}

Beyond the mechanism, the composition--temperature matrix is a selection chart for the phosphor engineer. The global optimum of the \ch{Mn^2+} emission, $x = 0.65$ annealed at $1100~^{\circ}$C, is reached at a smaller zinc surplus than the optimum we found for the \ch{Cr^3+}-doped analogue \cite{Rudolf2022}, in line with the different site and valence of the two activators. Near-stoichiometric crystals annealed at only $400~^{\circ}$C reach about half of this maximum under X-ray excitation, a bright phosphor at a fraction of the thermal energy required and, because their defect chemistry is still intact, a natural starting point for persistent-luminescence applications \cite{Uheda1997,Luchechko2019}, which we address in a separate study. The $580$~nm defect band, the red \ch{Cr^3+} emission and the \ch{Mn^4+}-assigned component each have their own optimum in the $(x, T_{\rm A})$ plane. The relative weights of these channels, and hence the emission colour under a given excitation, can therefore be set by the choice of $(x, T_{\rm A})$ alone, without changing the dopant (Fig.~\ref{fig:allin}).

\section{Conclusions}
\label{sec:conclusions}

Solution stoichiometry and annealing temperature act on different microscopic variables in Mn-doped \ch{ZnGa2O4} nanocrystals. The Zn/Ga ratio of the growth medium determines how much manganese is captured on tetrahedral sites and how much non-stoichiometry is stored as antisite and vacancy defects in the single-phase as-grown crystal. Annealing then transforms both inventories. At $400$--$600~^{\circ}$C, dehydroxylation and local ordering activate \ch{Mn^2+}, and the remaining defect population sensitises host-to-\ch{Mn^2+} energy transfer. Around $800~^{\circ}$C, native defects that are being annealed out still stabilise air-oxidised manganese, so that the \ch{Mn^2+} population and the X-ray- and $365$~nm-excited emission decrease. At $1100~^{\circ}$C, exsolution of $\beta$-\ch{Ga2O3}, relaxation of the lattice parameter to its stoichiometric value and recrystallisation into $100$--$270$~nm domains remove the compensating defects; divalent manganese is restored, and the green emission reaches its global maximum in the Zn-excess compositions whose final crystal is closest to ideal \ch{ZnGa2O4}. At the Zn-rich extreme, the dopant is rejected during growth, and no annealing treatment recovers the emission. The practical consequence is that the precursor ratio of a phosphor or scintillator synthesis should be chosen for the composition the crystal will reach after processing, not for the nominal formula.

Two further results follow from the same matrix. First, the additive-free hydrothermal route delivers single-crystalline nanocrystals whose habit is set by the solution ratio alone, from $\{111\}$-faceted octahedra under Ga excess to isotropic polyhedra under Zn excess. Second, the channel-resolved maps constitute a design chart. For X-ray scintillation and transducer applications, $x = 0.65$ annealed at $1100~^{\circ}$C gives the brightest \ch{Mn^2+} emission; for phosphors excited in the near-UV ($365$~nm), $x = 0.50$--$0.55$ at $1100~^{\circ}$C; and where the thermal budget or the defect inventory matters, for low-temperature processing or for persistent luminescence, $x = 0.45$--$0.50$ annealed at $400~^{\circ}$C retains about half of the maximum X-ray-excited output. Because the defect band, the \ch{Cr^3+} lines and the \ch{Mn^4+} component each have their own optimum, the emission colour under a given excitation can be tuned within the same host by moving along the map. The persistent luminescence of the low-temperature samples, and the trap structure behind it, exceed the scope of this work and will be reported separately.

\section{Methods}
\label{sec:methods}

\textbf{Synthesis.} Mn-doped \ch{ZnGa2O4} (ZGO:Mn, denoted ZGM) nanocrystals were synthesised hydrothermally following our previously reported procedure \cite{Beke2021,Rudolf2022}, with all precursor concentrations doubled. In a typical synthesis, $0.8$~mL of aqueous \ch{MnCl2 * 4 H2O} solution ($1$~mmol~L$^{-1}$) and $2.0$~mL of aqueous \ch{Ga(NO3)3 * $n$ H2O} solution ($2$~mol~L$^{-1}$, $99.9\%$) were combined to fix the Mn(II) dose and the Ga(III) content, and the Zn/Ga ratio of the series was adjusted through the added volume of \ch{Zn(NO3)2 * 6 H2O} stock solution ($2$~mol~L$^{-1}$, $99.9\%$): $0.6$, $0.9$, $1.0$, $1.1$, $1.3$ and $1.5$~mL, corresponding to nominal Zn/Ga $= 0.30$, $0.45$, $0.50$, $0.55$, $0.65$ and $0.75$ ($0.8$ and $1.2$~mL were used additionally for the reproducibility series). Each mixture was diluted to a total volume of $15$~mL with deionised water (Millipore Type~I) and stirred for $30$~min; approximately $2$~mL of ammonium hydroxide ($28\%$, Avantor--VWR, a.r.) was then added to reach pH~9, and stirring was continued for a further $30$~min. The precursor suspensions were sealed in a PTFE-lined autoclave (Berghof DAB-2) and held at $230~^{\circ}$C for $13$~h (Berghof DAH heating block with BTC-3000 temperature controller). The white products were collected by centrifugation, washed with $0.1$~mol~L$^{-1}$ HCl in isopropanol ($1{:}10$ v/v) and subsequently with isopropanol, and dried above $60~^{\circ}$C.

\textbf{Post-annealing.} The dried products were annealed either as loose powders or as pellets, following our previously reported protocol \cite{Beke2021}. Pellets of identical mass ($130$~mg) were pressed in a $13$~mm Specac die at $2.0$--$2.5$~t. Annealing was performed in air at $400$, $600$, $800$ and $1100~^{\circ}$C --- with an additional $300~^{\circ}$C step for the EPR series --- in ceramic crucibles or in quartz EPR ampoules, using either a box furnace (chamber $\approx20\times20\times30$~cm) or a tube furnace (length $30$~cm, diameter $5$~cm). All treatments included an intermediate hold at $350~^{\circ}$C; the heating rate was $3~^{\circ}$C~min$^{-1}$, and cooling in the box furnace took $4$--$6$~h depending on the annealing temperature.

\textbf{Powder X-ray diffraction.} XRD patterns were recorded on a Huber G670 Guinier imaging-plate camera using Cu~K$\alpha_{1}$ radiation ($\lambda = 1.54056$~\AA) from a focusing Johansson-type monochromator. The samples were measured at room temperature in glass capillaries of $0.5$~mm diameter and $0.01$~mm wall thickness. Rietveld refinements were performed with TOPAS-Academic \cite{Coelho2018}; refined lattice parameters, phase fractions and domain sizes, together with the treatment of instrumental broadening, are compiled in the Supporting Information. The sensitivity of the refinement to cation-distribution parameters was audited by comparing twelve nested parametrisations, releasing in all combinations the isotropic displacement parameters, the site occupancies, the spinel inversion degree and a Zn$\leftrightarrow$Ga exchange. Because \ch{Zn^2+} and \ch{Ga^3+} differ by one electron, the inversion- and exchange-type parameters are statistically indeterminate on these data: released alone they leave $R_{\rm wp}$ unchanged, and in the fully free model they refine to non-monotonic, partly bound-pinned values, while the isotropic displacement parameters alone reproduce the entire improvement of the fully free model. Final values are therefore quoted from the fixed-distribution parametrisation (unit occupancies, zero inversion, per-sample $B_{\rm eq}$).

\textbf{X-ray-excited optical luminescence.} XEOL was excited with the unfiltered, non-monochromatised radiation of a conventional sealed-tube X-ray source with Mo anode, comprising the Mo characteristic lines (K$\alpha$, $17.5$~keV; K$\beta$, $19.6$~keV) and bremsstrahlung up to $50$~keV; the Be exit window and the $90$~mm air path to the sample effectively suppress photon energies below $3$~keV. The powder samples were mounted on a holder with $8.5$~mm apertures, oriented at $45^{\circ}$ to the incident beam, and irradiated over a spot of $80$~mm$^2$. The emitted light was collected at $90^{\circ}$ to the excitation beam by an optical fibre (Ocean Optics QP600-025-UV-BX, NA $= 0.22$) placed $30$~mm from the sample surface and analysed with an Ocean Optics QE~Pro-FL spectrometer ($350$--$1100$~nm; optical resolution $\approx6.4$~nm with the HC1 grating and a $200$~\textmu m slit). For the identification of zero-phonon lines, an Ocean Optics QE65000 spectrometer with $0.25$~nm resolution was used.

\textbf{Photoluminescence and excitation spectroscopy.} PL and PLE spectra were recorded on a Horiba Jobin-Yvon NanoLog FL3-2iHR spectrofluorometer equipped with a $450$~W xenon lamp, an iHR-320 grating spectrometer and an R928P photomultiplier tube. Colloidal samples were measured in $10$~mm quartz cuvettes (Hellma 110-QS). PLE spectra were recorded at an emission wavelength of $695$~nm with $10$~nm bandpass. Channel-resolved intensities (Fig.~\ref{fig:channels}) were obtained by decomposing each spectrum into the \ch{Mn^2+}, $580$~nm, \ch{Cr^3+} and (under $365$~nm excitation) \ch{Mn^4+}-assigned components. For large-area UV illumination a $400$~W lamp was used.

\textbf{Electron paramagnetic resonance.} X-band EPR spectra ($\approx9.4$~GHz, resonance field $\approx0.34$~T) were recorded at room temperature on a Magnettech MiniScope MS-400 spectrometer with a TE$_{102}$ rectangular cavity. A continuous flow of dry nitrogen prevented moisture condensation in the cavity. The samples were contained in high-quality, defect-free quartz ampoules (Wilmad-LabGlass). A low microwave power ($2$~mW) and low field-modulation amplitude ($0.10$--$0.25$~mT) were employed to avoid distortion of the line shapes. Because the measurements were carried out at room temperature, the charge states of the defects correspond to thermal equilibrium under ambient conditions. \ch{Mn^2+} contents were compared through the double integral of the sextet, normalised to the sample mass.

\textbf{Scanning electron microscopy and EDS.} SEM images and energy-dispersive X-ray spectra were recorded on a TESCAN MIRA3 microscope equipped with an Element EDS system, using the In-Beam secondary-electron detector. For the measurements, the nanoparticles were dispersed in water and drop-cast onto Si substrates.

\textbf{Raman spectroscopy.} Raman spectra were acquired with a Renishaw inVia confocal Raman microscope using $532$~nm excitation ($2$~W continuous-wave source operated at $1\%$; approximately $20$~mW at the sample). For the composition--annealing maps (Fig.~\ref{fig:S1}), the $A_{1g}$ ($\approx714$~cm$^{-1}$) and $F_{2g(3)}$ ($\approx611$~cm$^{-1}$) bands were fitted and their integrated areas and inverse linewidths (sharpness, $1/\mathrm{FWHM}$) are mapped over the $(x, T_{\rm A})$ plane; the linewidth serves as a normalisation-free order parameter, whereas absolute intensities depend on scattering volume, particle size, aggregation and focusing and are not compared quantitatively between samples.

\begin{acknowledgments}

This work was supported by the Doctoral Excellence Fellowship Programme (DCEP), funded by the National Research, Development and Innovation Fund of the Ministry of Culture and Innovation and by the Budapest University of Technology and Economics. The Hungarian research infrastructure was provided by the Hungarian Academy of Sciences. The research was funded by project no.~TKP2021-NVA-04 of the Ministry of Innovation and Technology of Hungary, financed from the National Research, Development and Innovation Fund (NKFIH) under the TKP2021-NVA funding scheme, and by the Quantum Information National Laboratory sponsored by the Ministry of Innovation and Technology of Hungary via the NKFIH. This work was also supported by the National Research, Development and Innovation Office of Hungary (NKFIH) and by the Ministry of Culture and Innovation (Grant No.~149457).
\end{acknowledgments}

\bibliographystyle{apsrev4-2}
\bibliography{references}

\clearpage
\onecolumngrid
\setcounter{figure}{0}
\setcounter{table}{0}
\renewcommand{\thefigure}{S\arabic{figure}}
\renewcommand{\thetable}{S\arabic{table}}

\section*{Supporting Information}

\begin{figure}[htp]
    \includegraphics[width=\columnwidth]{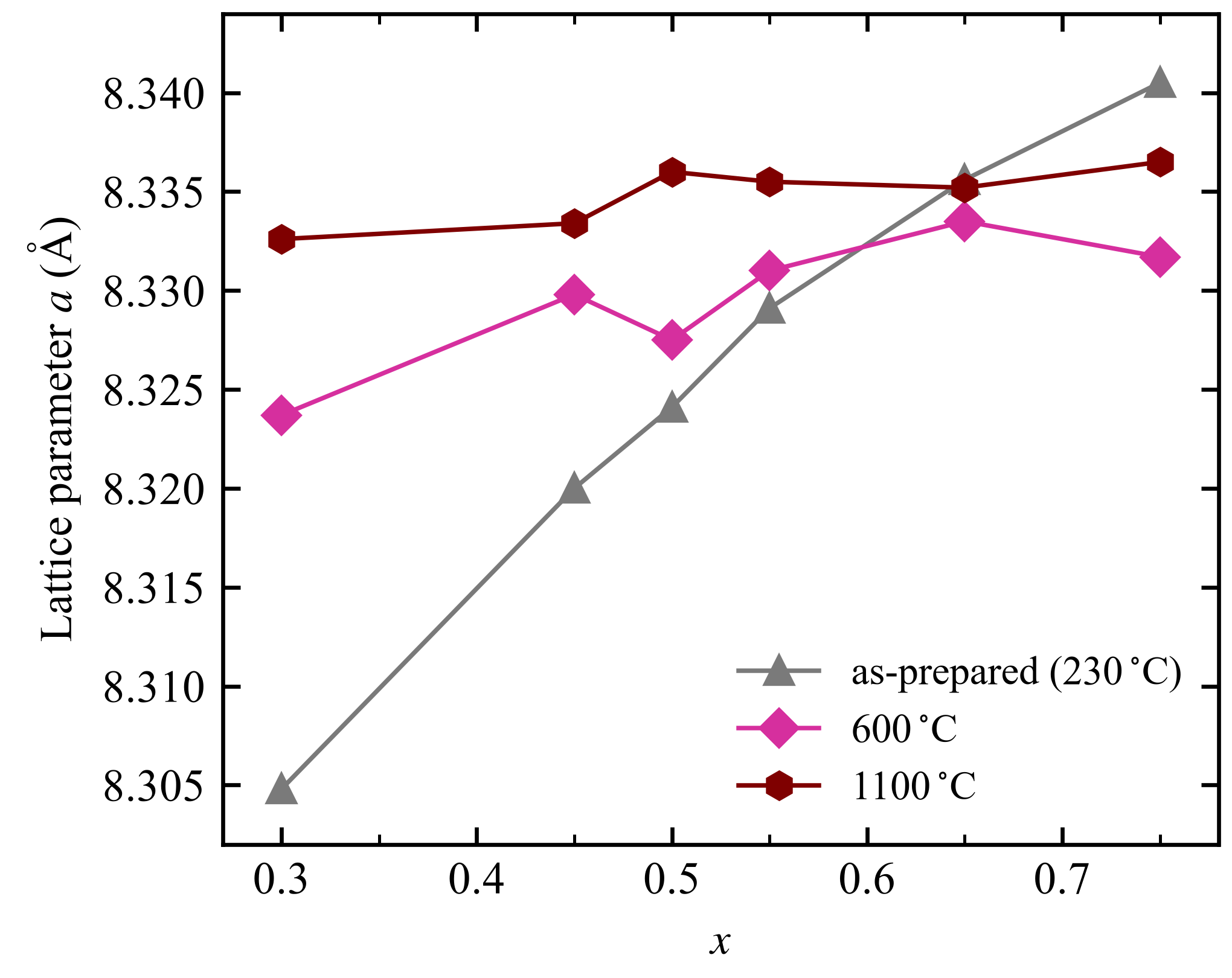}
    \caption{
        Refined cubic lattice parameter $a$ as a function of $x$ for the as-prepared samples ($230~^{\circ}$C) and after annealing at $600$ and $1100~^{\circ}$C. }
    \label{fig:S26}
\end{figure}

\begin{figure*}[htp]
    \includegraphics[width=\textwidth]{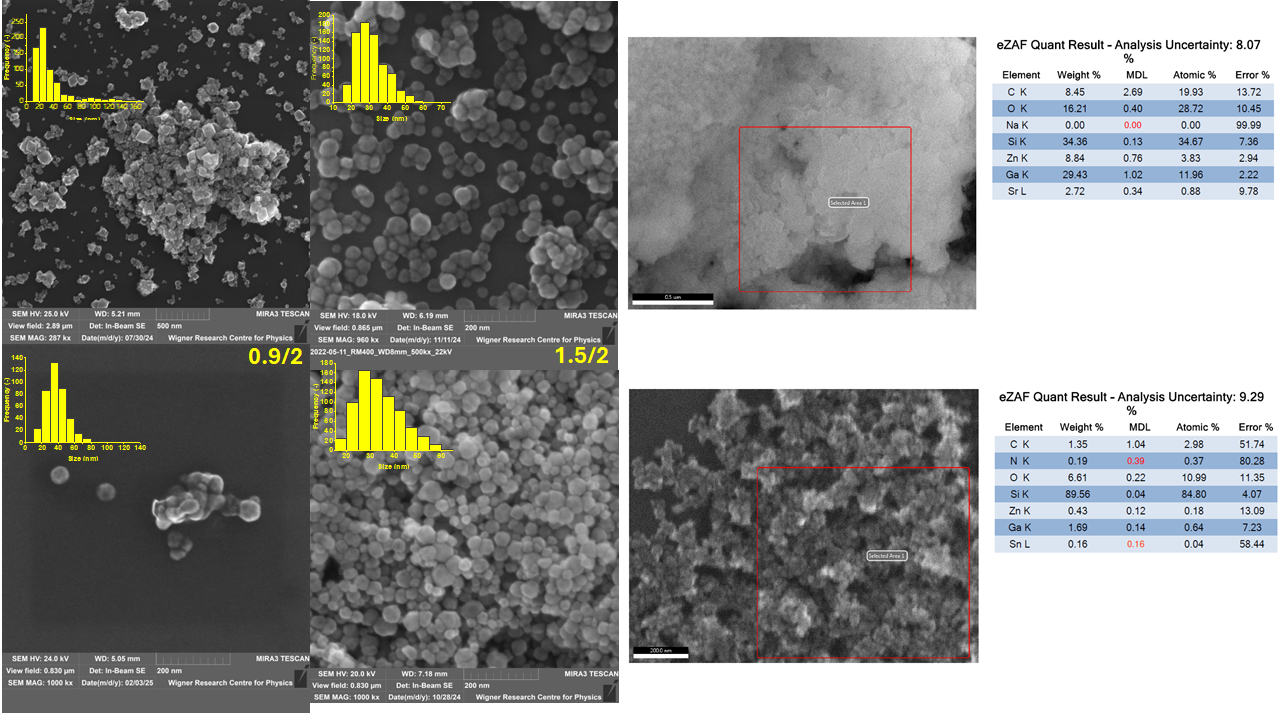}
    \caption{
        Additional SEM images of the $x = 0.45$ and $0.75$ samples with size histograms, and representative SEM-EDS quantification tables (Si substrate signal included).}
    \label{fig:S3}
\end{figure*}
\clearpage

\begin{figure*}[htp]
    \includegraphics[width=0.85\textwidth]{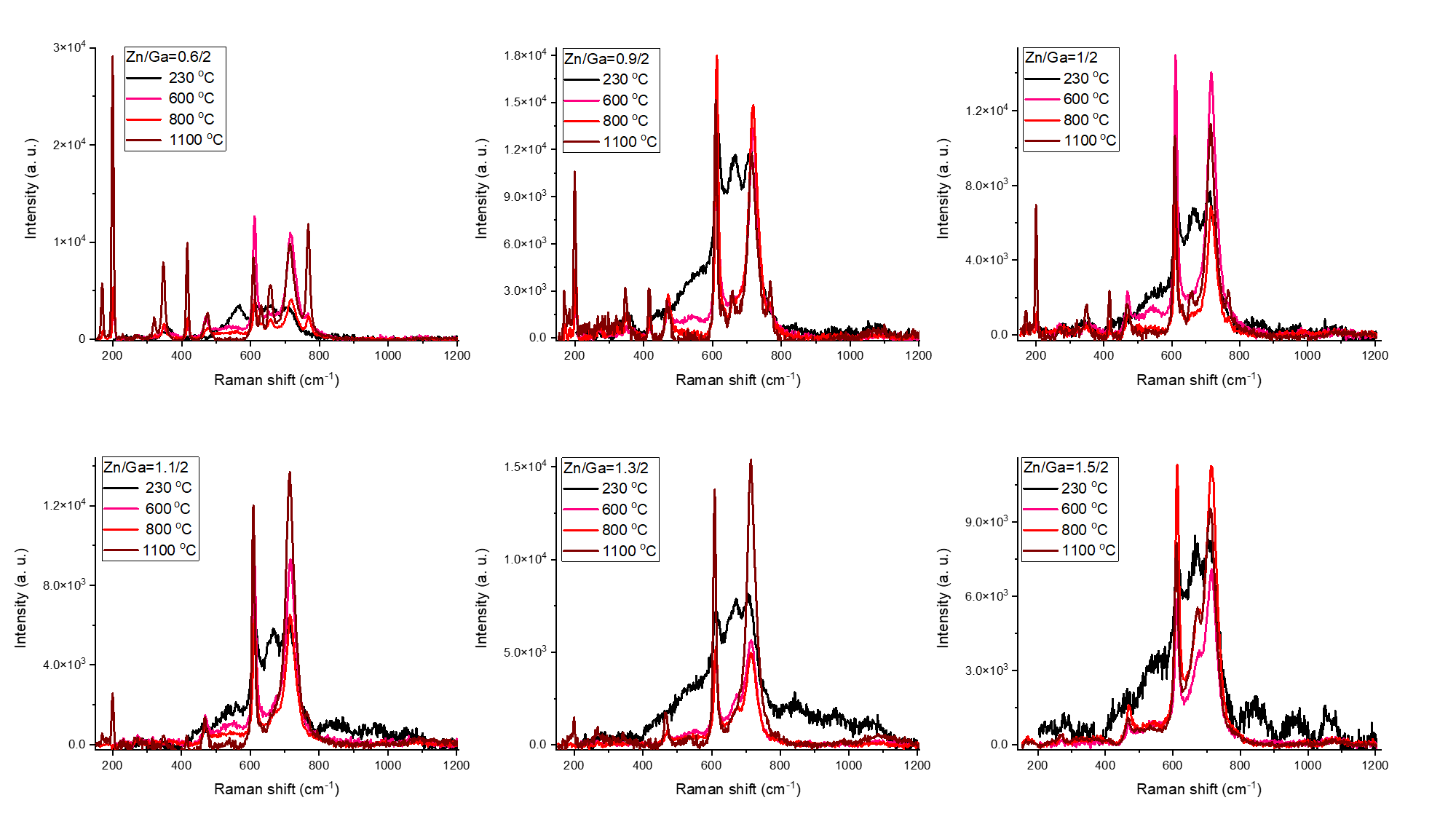}
    \caption{
        Raman spectra of each composition after the $230$, $600$, $800$ and $1100~^{\circ}$C treatments.}
    \label{fig:S5}
\end{figure*}
\clearpage

\begin{figure*}[htp]
    \includegraphics[width=\textwidth]{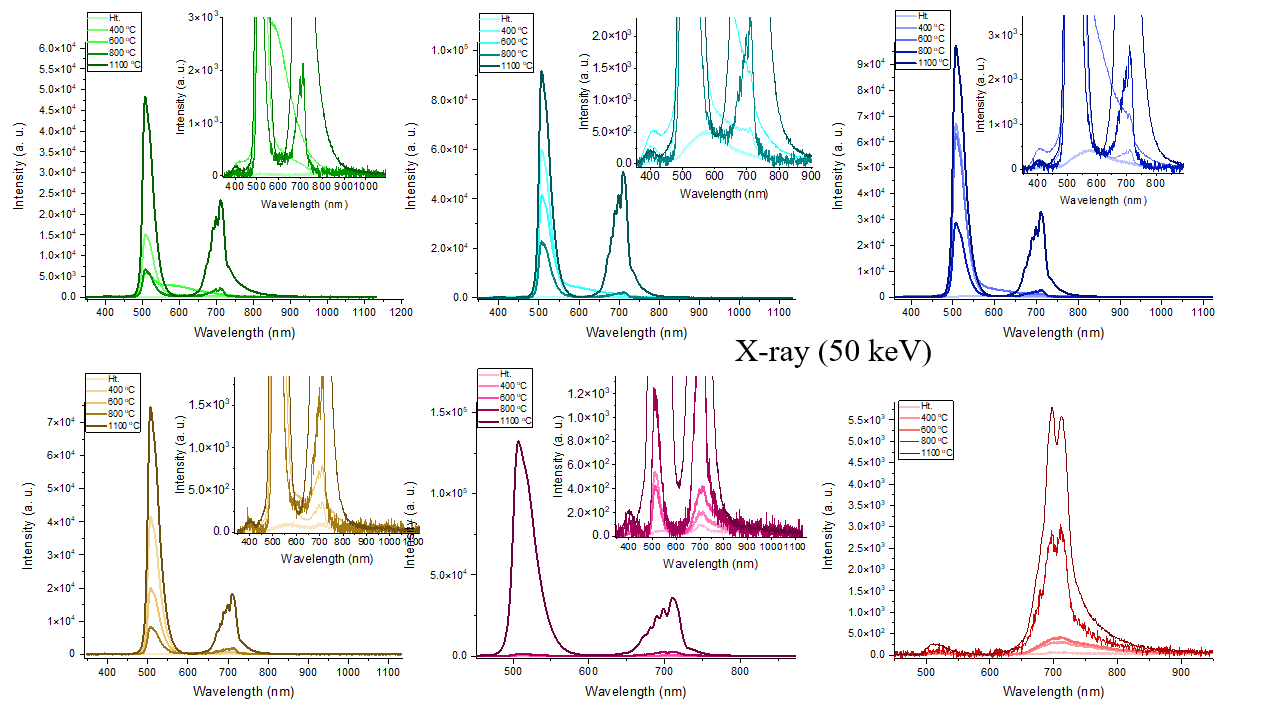}
    \caption{
        XEOL spectra of all compositions at each annealing stage; insets: magnified views showing the \ch{Cr^3+} $688$/$696$/$704$--$715$~nm multiplet and the weak defect bands.}
    \label{fig:S6}
\end{figure*}
\clearpage

\begin{figure*}[htp]
    \includegraphics[width=0.8\textwidth]{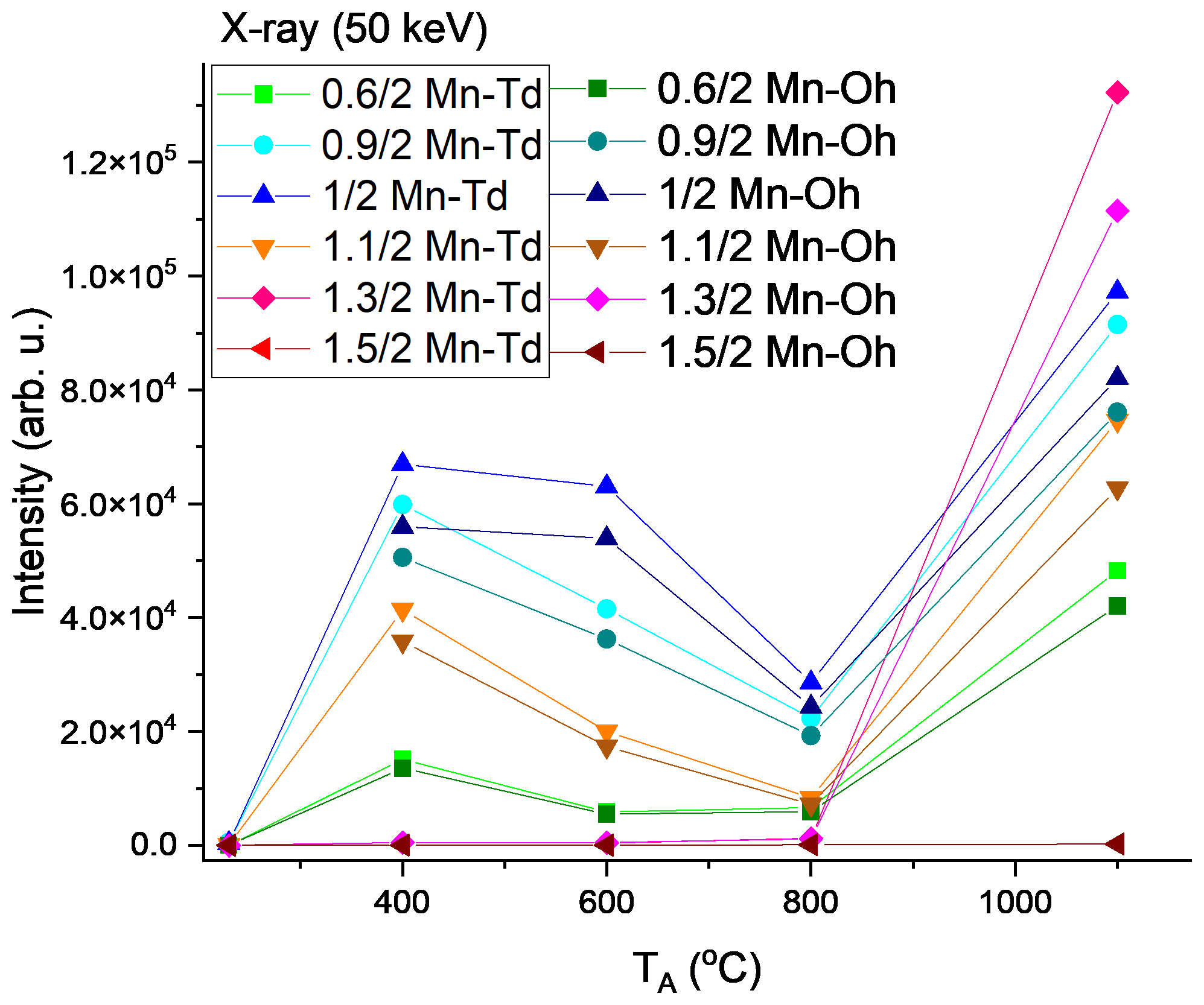}
    \caption{
        Decomposition of the XEOL green band into Mn-$T_{\rm d}$ and Mn-$O_{\rm h}$ components versus annealing temperature. }
    \label{fig:S12}
\end{figure*}
\clearpage

\begin{figure*}[htp]
    \includegraphics[width=\textwidth]{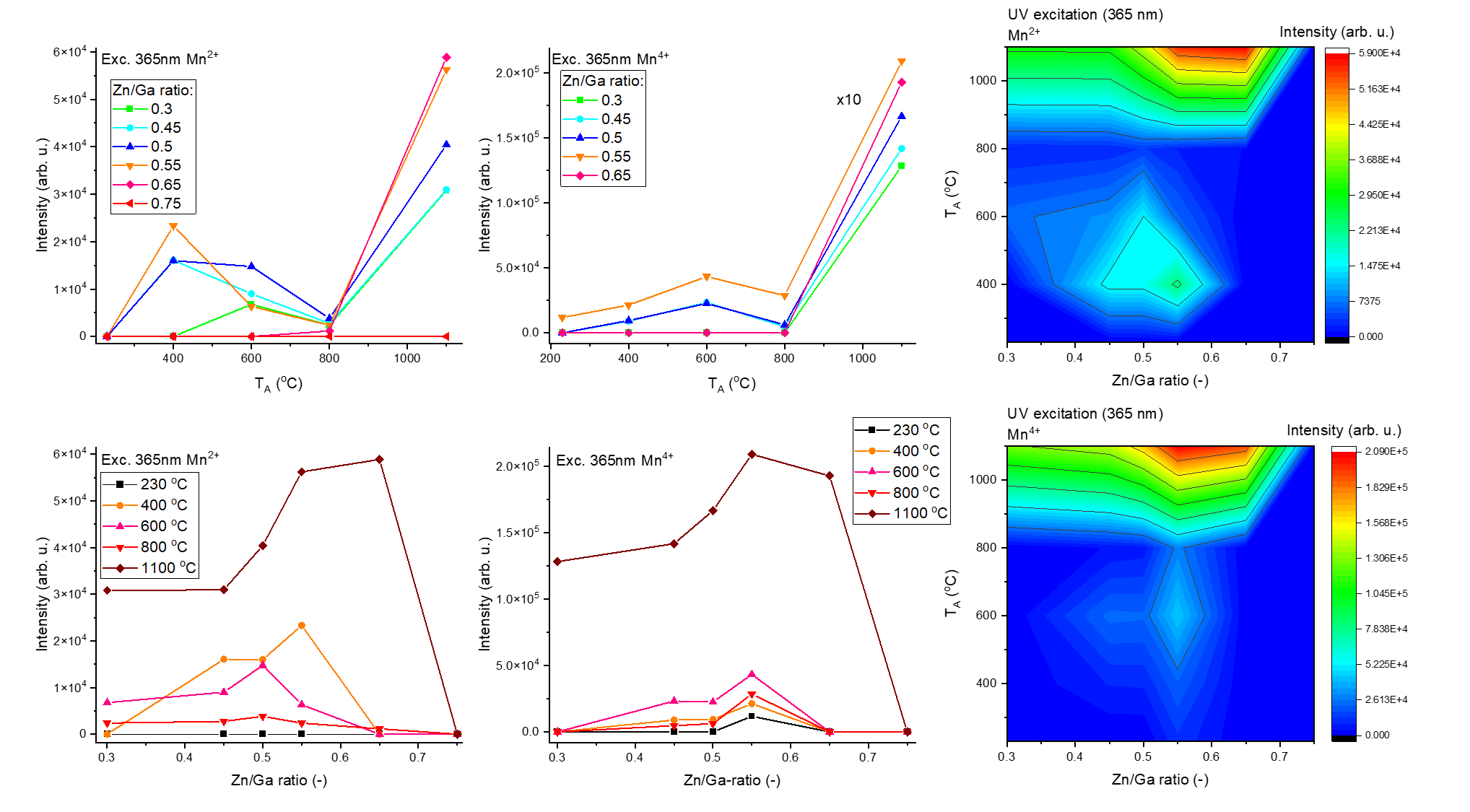}
    \caption{
        Emission of \ch{Mn^2+} (green band) and of the red component assigned to \ch{Mn^4+} under $365$~nm excitation versus annealing temperature (top) and composition (bottom), with the corresponding composition--annealing contour maps (right). The \ch{Mn^4+}-assigned component is confined to $x = 0.45$--$0.55$ up to $800~^{\circ}$C, becomes strong for all $x \leq 0.65$ after $1100~^{\circ}$C, and is absent at $x = 0.75$ at every annealing stage.}
    \label{fig:S2}
\end{figure*}
\clearpage

\begin{figure*}[htp]
    \includegraphics[width=\textwidth]{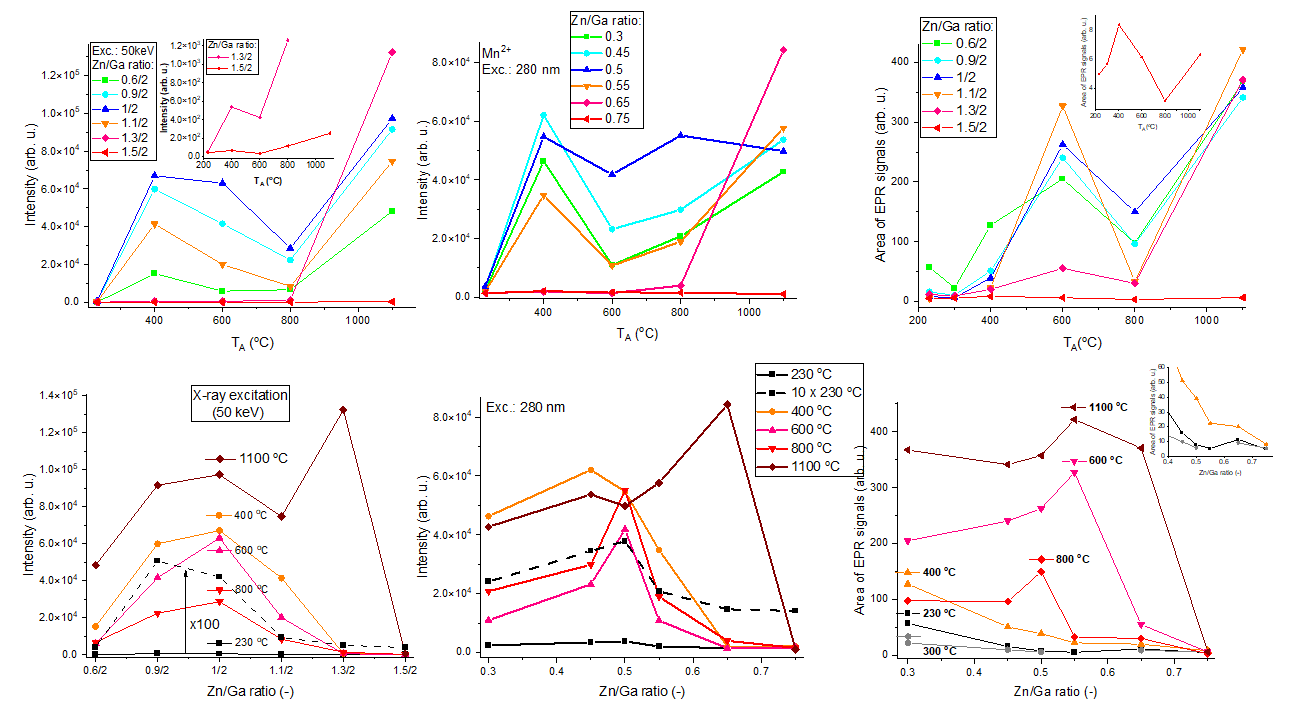}
    \caption{
        Line profiles of the composition--annealing data. Top: \ch{Mn^2+} emission under X-ray ($50$~keV, left) and $280$~nm (centre) excitation and \ch{Mn^2+} EPR double integral (right) versus annealing temperature for each composition. Bottom: the same quantities versus composition for each annealing temperature.}
    \label{fig:Sint}
\end{figure*}
\clearpage

\begin{figure*}[htp]
    \includegraphics[width=\textwidth]{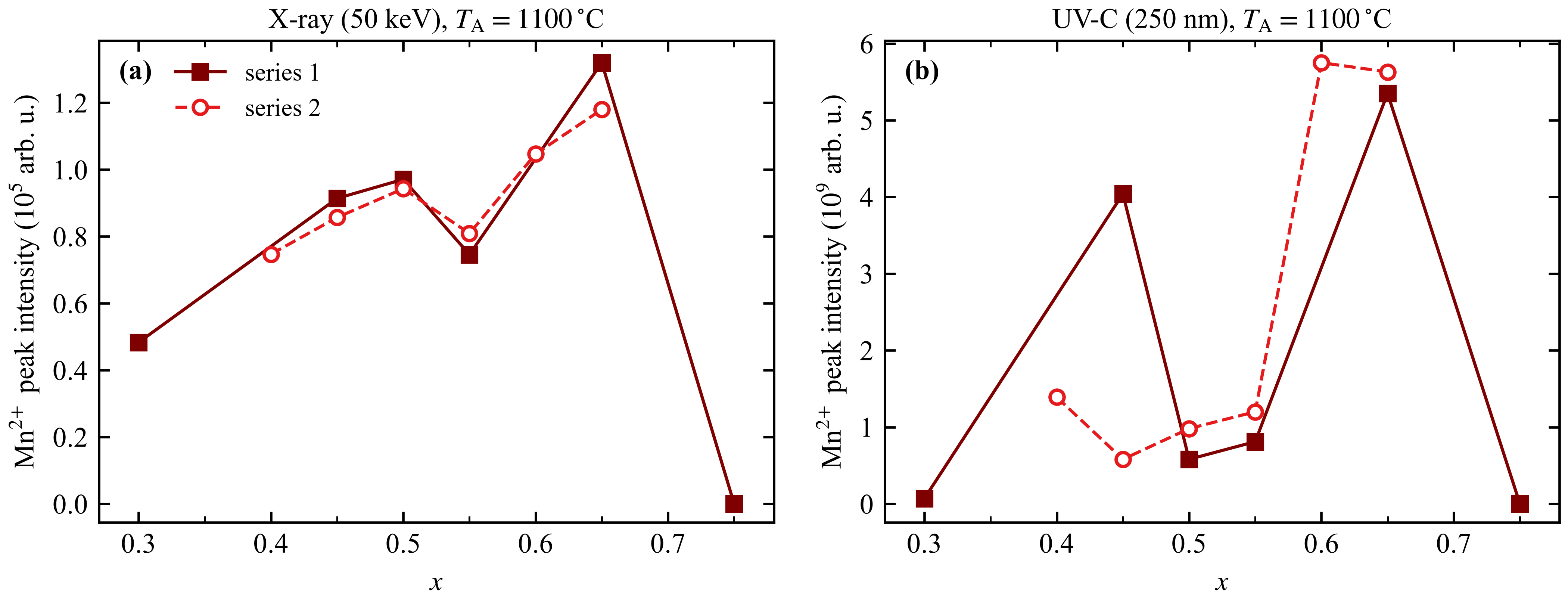}
    \caption{
        Reproducibility of the $1100~^{\circ}$C composition profile. Peak intensity of the \ch{Mn^2+} emission of two independently synthesised series under (a) X-ray ($50$~keV) and (b) UV-C ($250$~nm) excitation, as a function of $x$. Series~2 additionally includes $x = 0.40$ and $0.60$.}
    \label{fig:S7}
\end{figure*}
\clearpage

\begin{figure*}[htp]
    \includegraphics[width=\textwidth]{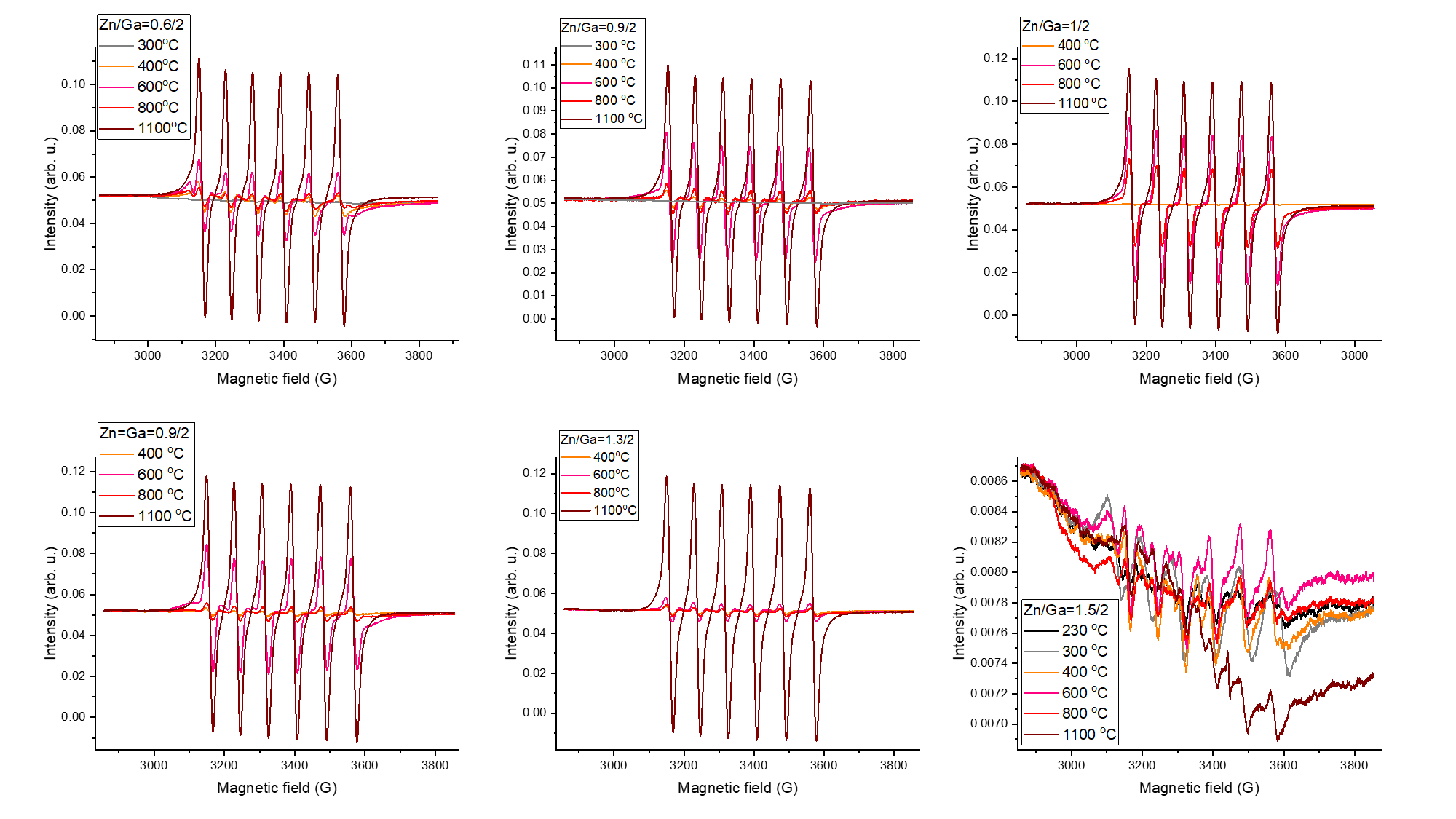}
    \caption{
        X-band EPR spectra of all compositions and annealing stages. Note the $\approx15\times$ smaller ordinate scale of the $x = 0.75$ panel.}
    \label{fig:S13}
\end{figure*}
\clearpage

\begin{figure*}[htp]
    \includegraphics[width=\textwidth]{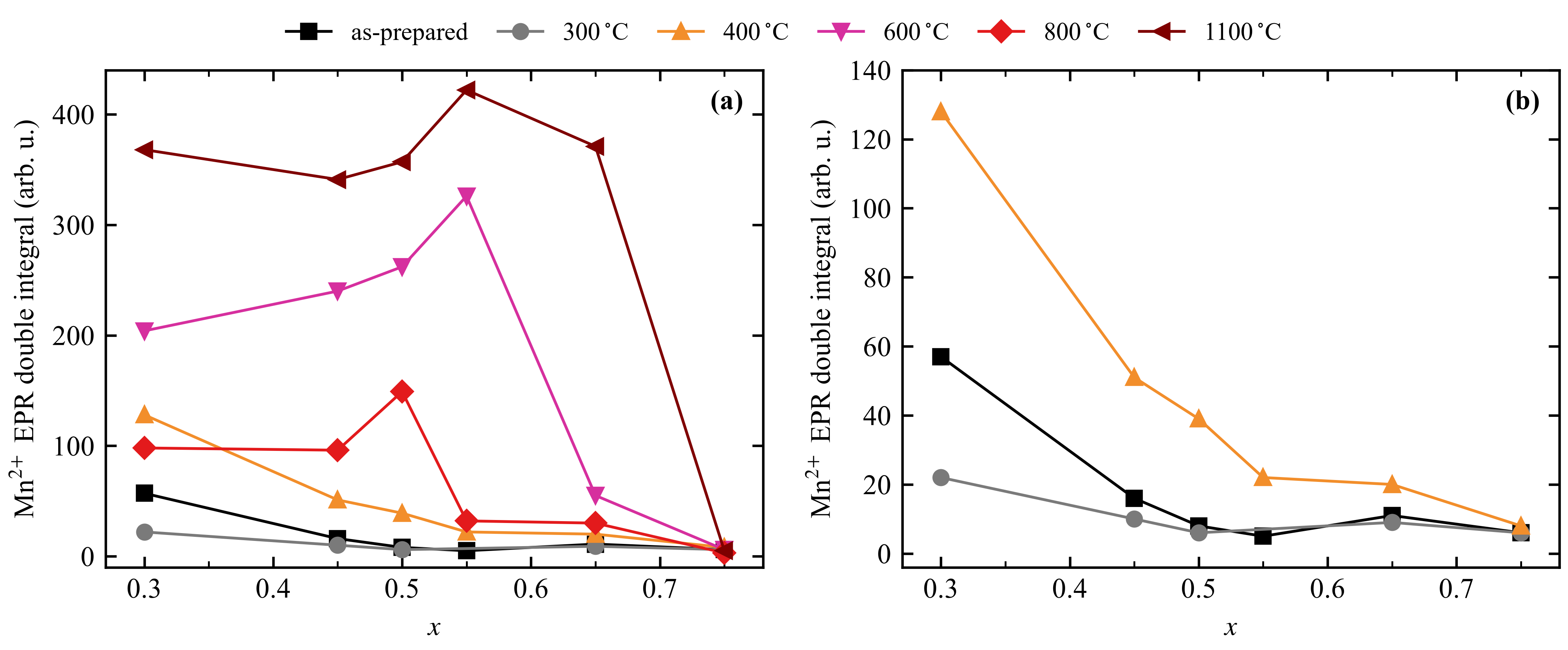}
    \caption{
        Double integral of the \ch{Mn^2+} EPR sextet as a function of $x$ for the as-prepared samples and after annealing at $300$--$1100~^{\circ}$C (a); (b) enlarged view of the as-prepared, $300$ and $400~^{\circ}$C series.}
    \label{fig:S14}
\end{figure*}
\clearpage

\begin{figure*}[htp]
    \includegraphics[width=0.9\textwidth]{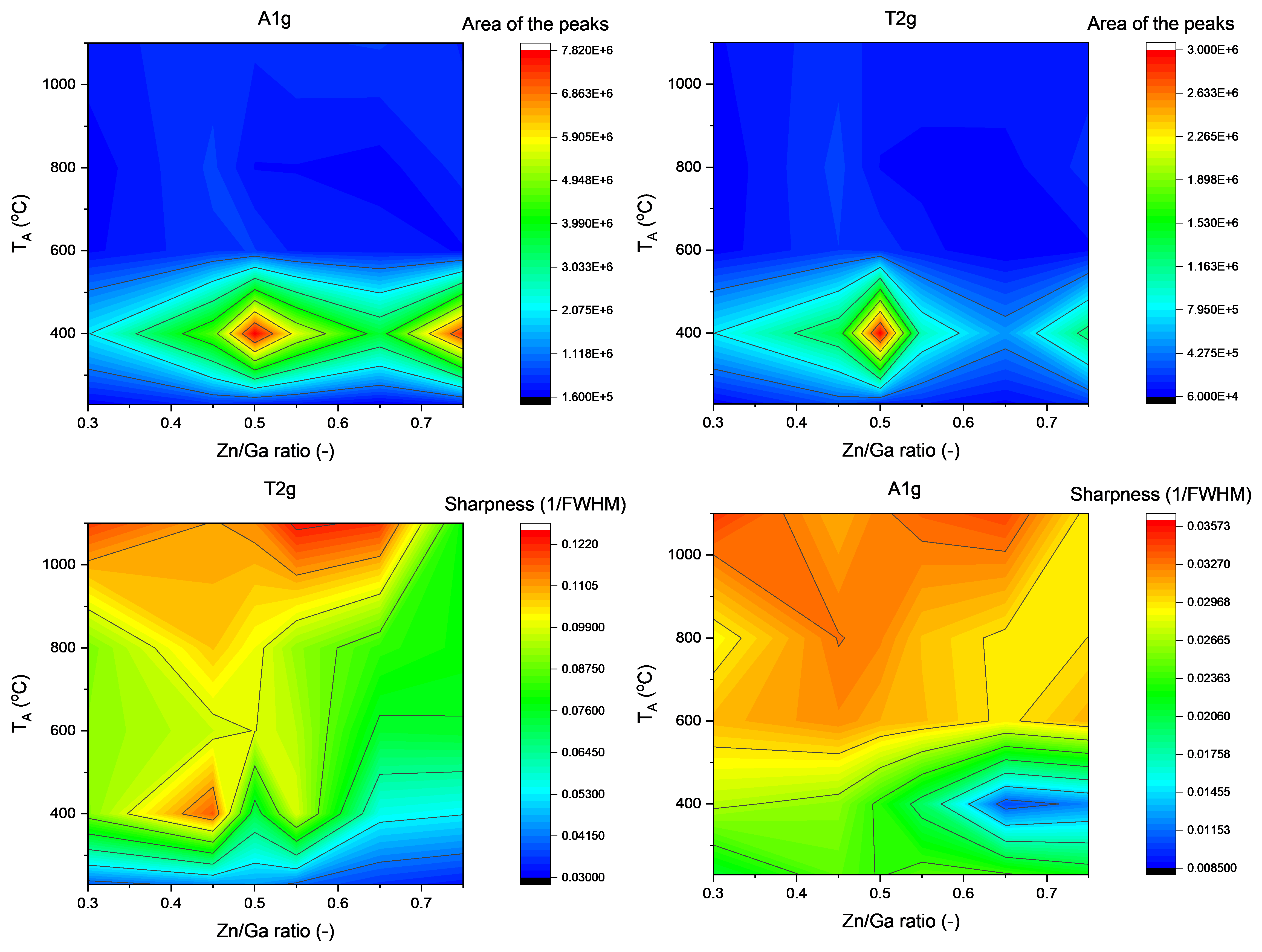}
    \caption{
        Composition--annealing maps of the fitted \ch{ZnGa2O4} Raman parameters: integrated peak area (upper pair: $A_{1g}$ left, $F_{2g(3)}$/$T_{2g}$ right) and sharpness, $1/\mathrm{FWHM}$ (lower pair: $T_{2g}$ left, $A_{1g}$ right). The area maps peak at $(x, T_{\rm A}) = (0.5, 400~^{\circ}\mathrm{C})$ with a secondary maximum at $x = 0.75$; the linewidth maps are the normalisation-free order parameters used in the main text: the $F_{2g(3)}$ sharpness mirrors the UV-excited \ch{Mn^2+} output (two maxima at $x = 0.45$ and $0.55$ at $400~^{\circ}$C, no minimum at $800~^{\circ}$C), whereas the $A_{1g}$ sharpness peaks at $x = 0.50$; neither shows an anomaly at $800~^{\circ}$C.}
    \label{fig:S1}
\end{figure*}
\clearpage

\begin{figure*}[htp]
    \includegraphics[width=\textwidth]{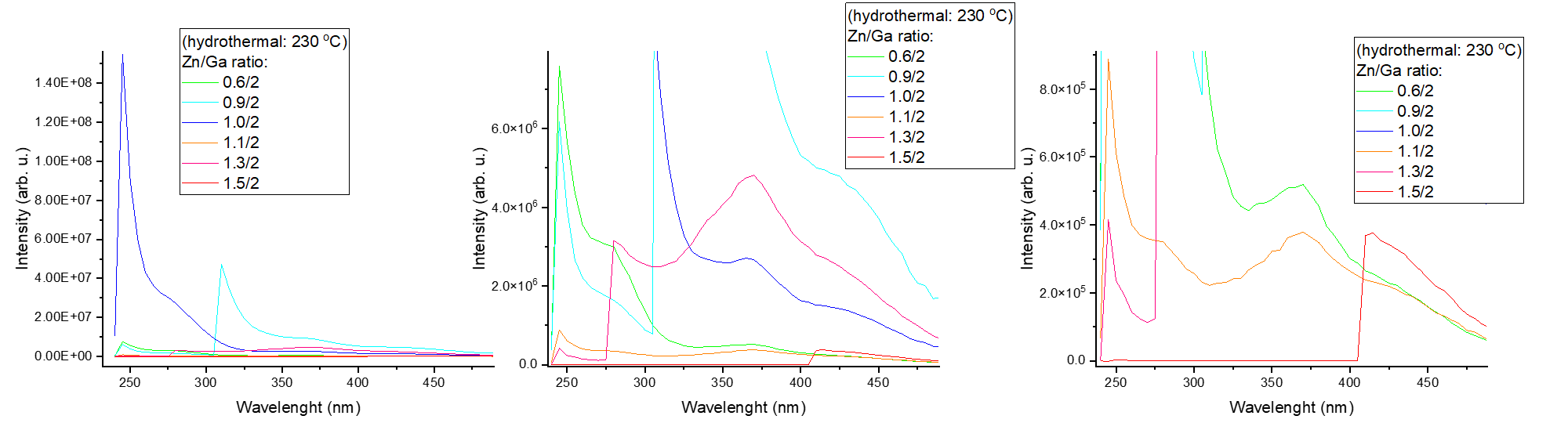}
    \caption{
        PLE spectra of the as-prepared samples (three intensity scales).}
    \label{fig:S22}
\end{figure*}
\clearpage

\begin{table}[htp]
{%
    \caption{Wavelength of the PLE band maximum (nm) versus composition and annealing temperature. }
    \label{tab:S1}
    \begin{tabular}{|c|c|c|c|c|c|c|}
    \hline
    $T_{\rm A}$ ($^{\circ}$C) $\backslash$ $x$ & $0.30$ & $0.45$ & $0.50$ & $0.55$ & $0.65$ & $0.75$ \\
    \hline
    $230$ & 245 & 355 & 245 & 245 & 370 & 415 \\
    $400$ & 250 & 250 & 250 & 250 & 430 & 250 \\
    $600$ & 290 & 245 & 240 & 245 & 245 & 365 \\
    $800$ & 245 & 245 & 245 & 245 & 245 & 245 \\
    $1100$ & 290 & 250 & 270 & 250 & 245 & 270 \\
    \hline
    \end{tabular}
}
\end{table}

\begin{figure*}[htp]
    \includegraphics[width=\textwidth]{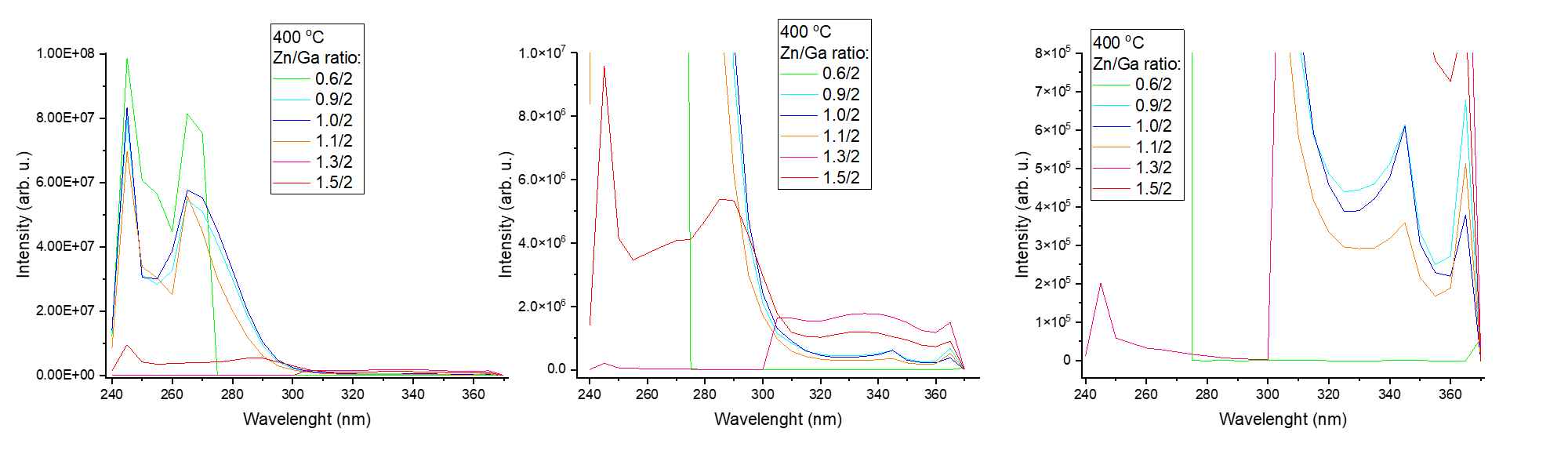}
    \caption{PLE spectra after $400~^{\circ}$C.}
    \label{fig:S23}
\end{figure*}
\clearpage

\begin{figure*}[htp]
    \includegraphics[width=\textwidth]{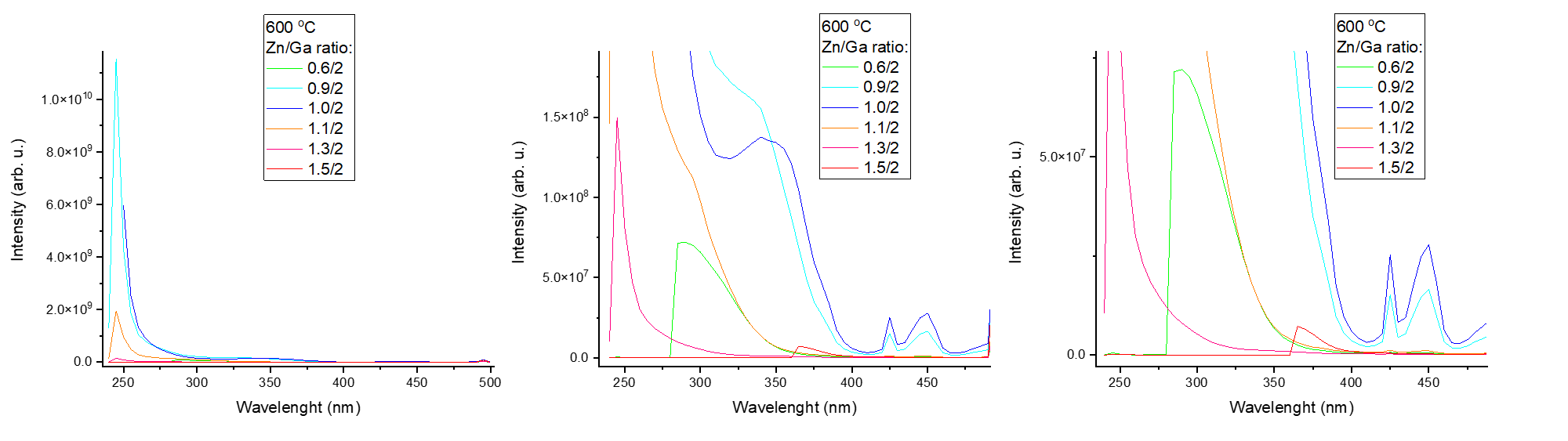}
    \caption{PLE spectra after $600~^{\circ}$C. }
    \label{fig:S24}
\end{figure*}
\clearpage

\begin{figure*}[htp]
    \includegraphics[width=0.85\textwidth]{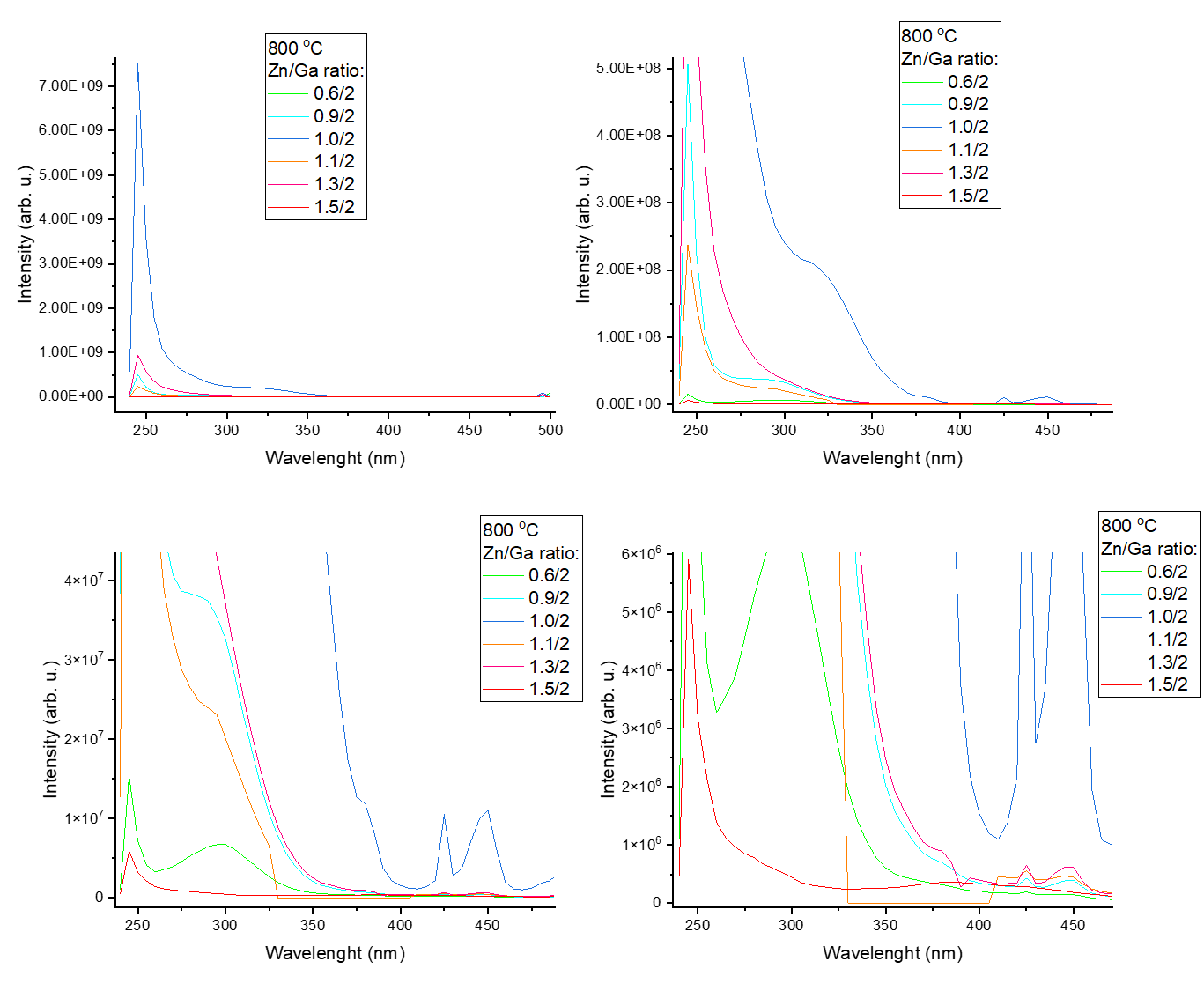}\\[6pt]
    \includegraphics[width=0.85\textwidth]{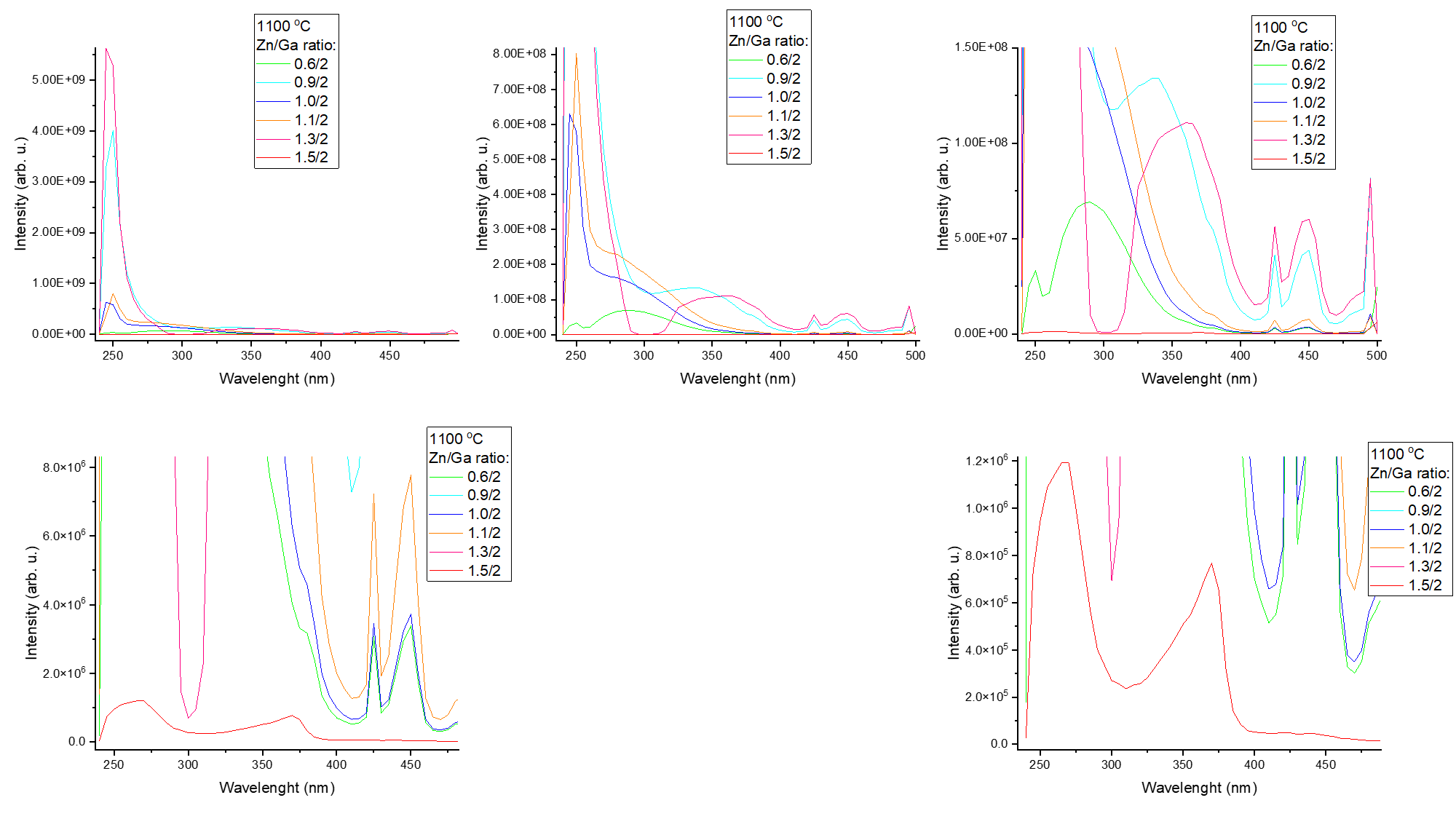}
    \caption{PLE spectra after $800$ and $1100~^{\circ}$C.}
    \label{fig:S25}
\end{figure*}
\clearpage

\begin{figure*}[htp]
    \includegraphics[width=0.7\textwidth]{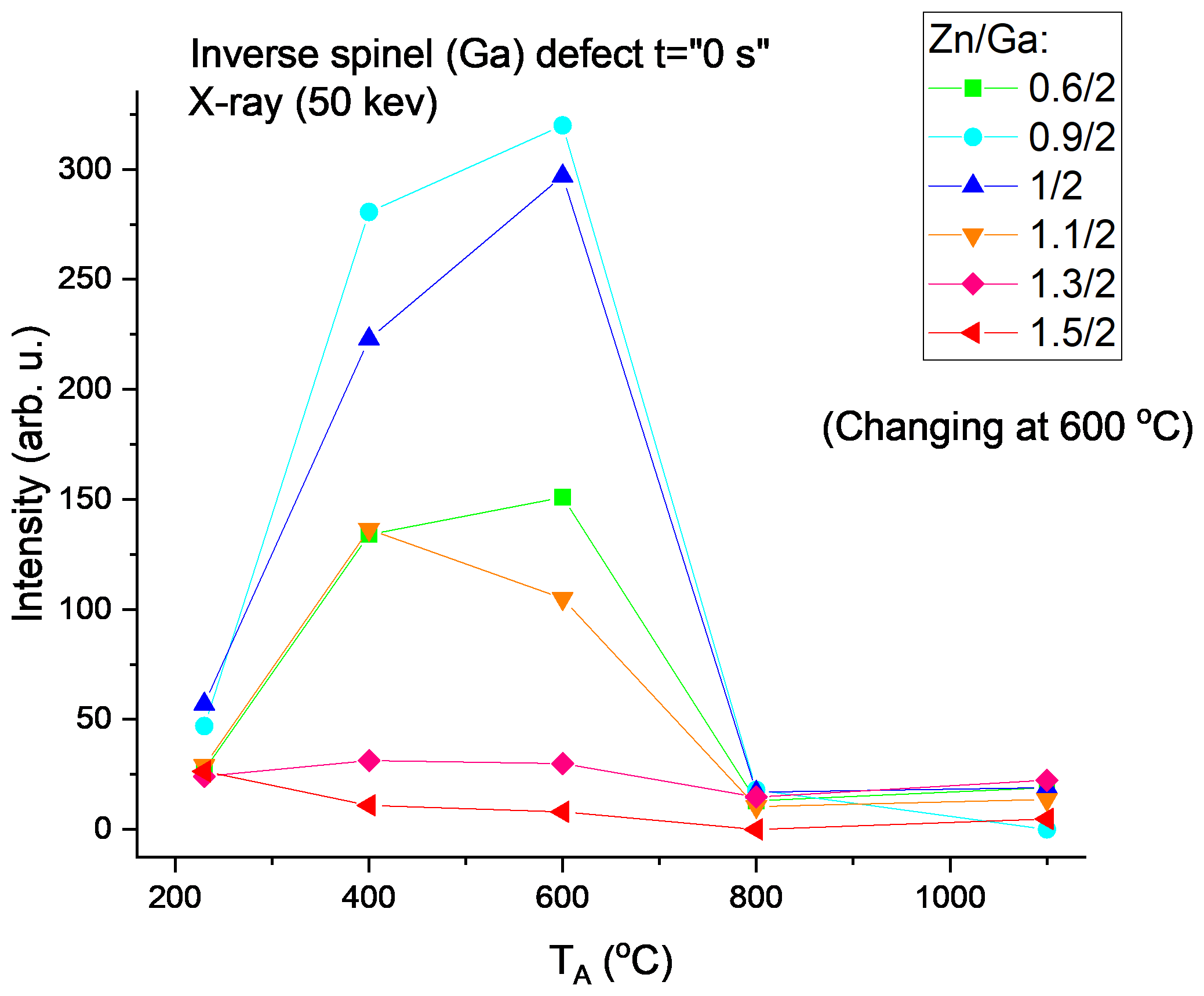}
    \caption{
        Defect-related XEOL component versus annealing temperature for each composition.}
    \label{fig:S8}
\end{figure*}
\clearpage

\begin{figure*}[htp]
    \includegraphics[width=0.8\textwidth]{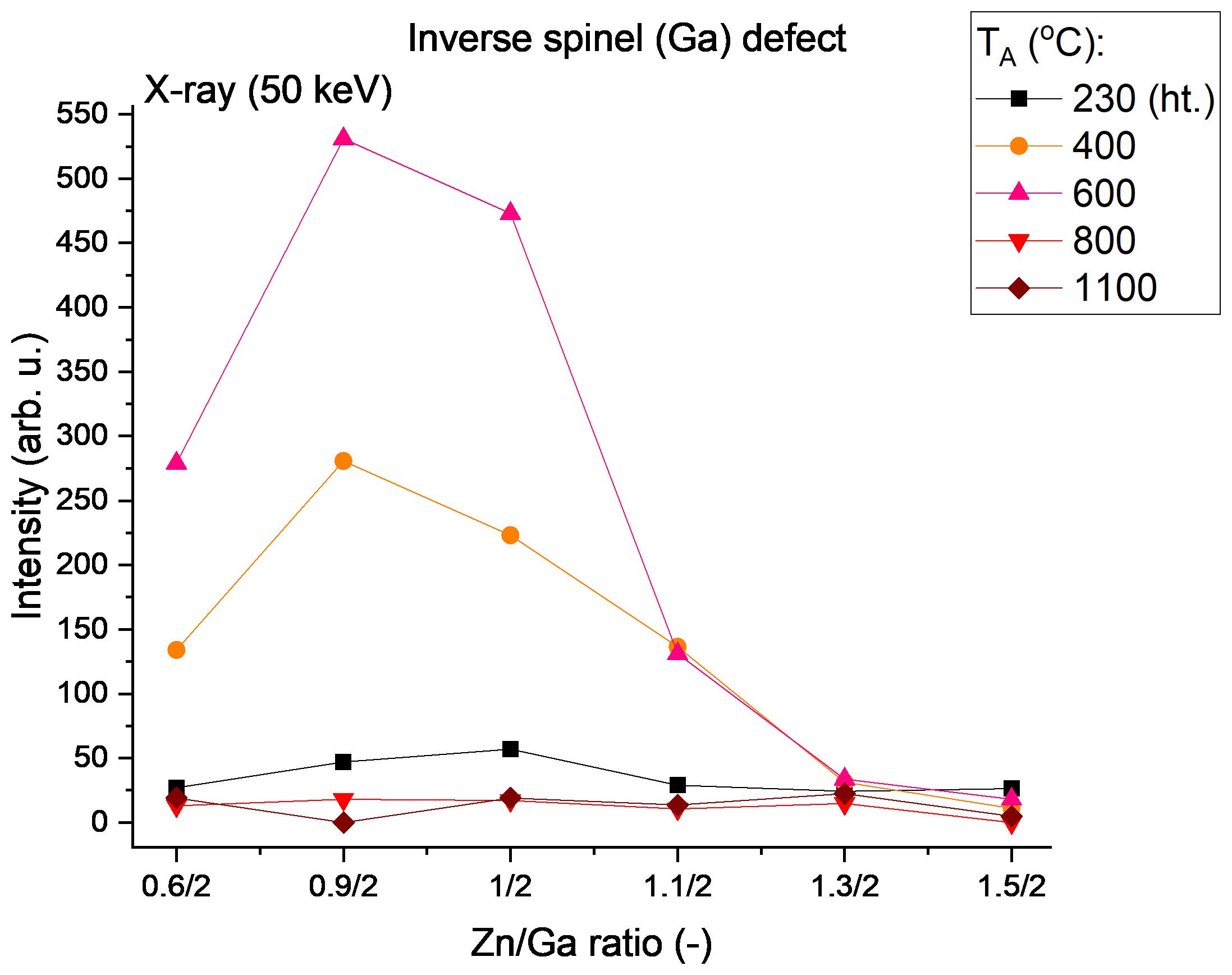}
    \caption{
        Defect-related XEOL component versus composition at each annealing temperature.}
    \label{fig:S9}
\end{figure*}
\clearpage

\begin{figure*}[htp]
    \includegraphics[width=\textwidth]{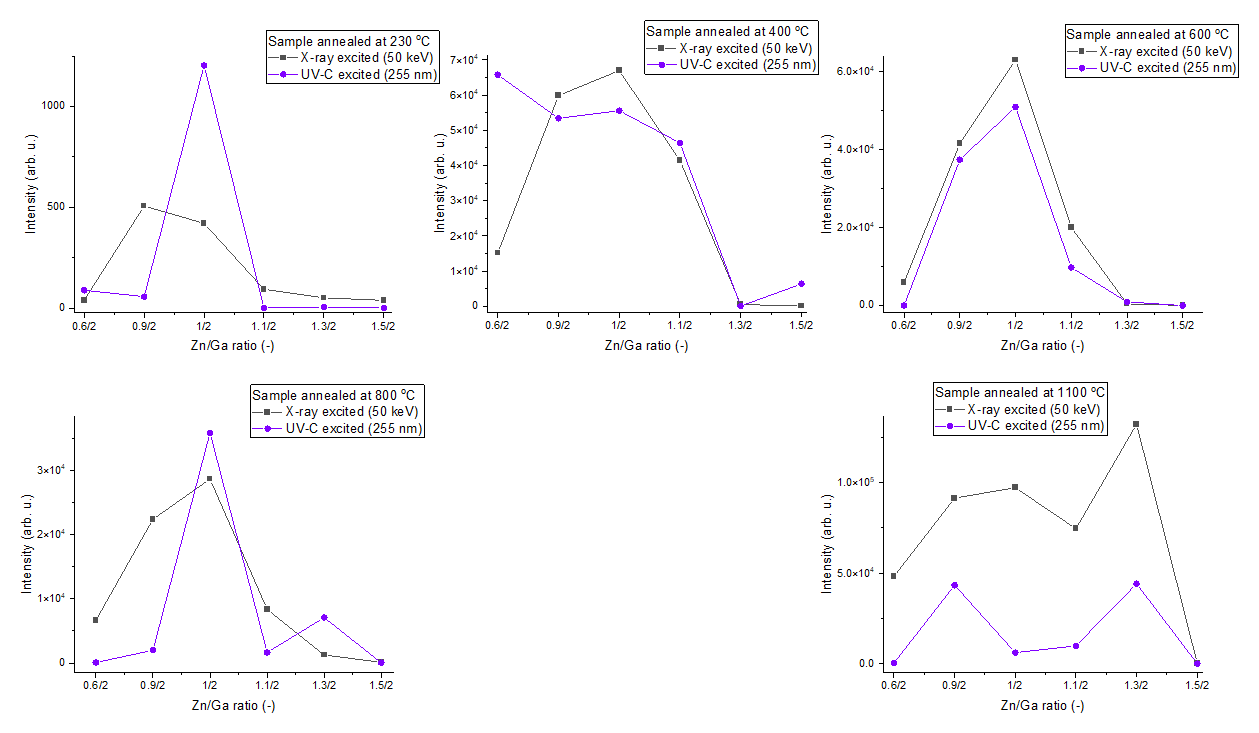}
    \caption{
        X-ray- versus UV-C-excited intensities as a function of composition at each annealing temperature.}
    \label{fig:S20}
\end{figure*}
\clearpage

\end{document}